\documentclass[12pt,a4paper,twoside,fleqn]{article}
\usepackage{color,graphicx,lscape,here,geometry,setspace,multirow,enumitem}
\usepackage[dvipsnames]{xcolor}
\usepackage{graphicx}
\usepackage{natbib}
\usepackage{longtable}
\usepackage{threeparttablex}
\usepackage{amsfonts}
\usepackage{authblk}
\usepackage{silence}

\usepackage[misc,geometry]{ifsym}

\usepackage{rotating}
\usepackage{multirow}

\usepackage{booktabs}
\usepackage{diagbox}
\usepackage{color,colortbl}
\usepackage{setspace}
\usepackage{algorithm2e}
\usepackage{enumitem}
\usepackage{amsfonts,amsmath,amssymb}
\usepackage{tikz}

\definecolor{nblue}{HTML}{000660}
\usepackage[colorlinks=true,urlcolor=nblue,linkcolor=nblue,citecolor=nblue]{hyperref}
\usepackage{bigstrut}

\newcommand{\KAdd}{\text{\texttt{A}}}
\newcommand{\KMul}{\text{\texttt{M}}}
\newcommand{\Wf}{\text{\texttt{W}}}

\newcommand{\Df}{\text{\texttt{D}}}
\newcommand{\Ef}{\text{\texttt{E}}}

\newcommand{\inlinehead}[1]{{\sffamily\textsc{\textbf{#1}}}. }

\usepackage{tabularx}
\usepackage{threeparttable}
\usepackage{dcolumn}
\newcolumntype{d}[1]{D{.}{.}{#1}}

\usepackage{array}
\newcolumntype{C}[1]{>{\centering\arraybackslash}p{#1}}

\usepackage{etoolbox} 
\makeatletter
\patchcmd{\BR@backref}{\newblock}{\newblock[}{}{}
\patchcmd{\BR@backref}{\par}{]\par}{}{}
\makeatother

\usepackage{pdflscape}
\usepackage{afterpage}

\usepackage{longtable}
\usepackage{array}

\usepackage{comment}
\usepackage{mathtools}
\usepackage[hang,flushmargin]{footmisc} 
\usepackage[hang]{footmisc}
\usepackage[british]{babel}

\pdfoutput=1
\usepackage{float}
\restylefloat{table}

\usepackage[title,titletoc]{appendix}
\makeatletter

\renewenvironment{appendices}{%
    \begin{oldappendices}%
    \renewcommand{\thefigure}{\ifnum \c@section>\z@ \thesection.\fi\@arabic\c@figure}%
    \@addtoreset{figure}{section}%
    \renewcommand{\thetable}{\ifnum \c@section>\z@ \thesection.\fi\@arabic\c@table}%
    \@addtoreset{table}{section}}{%
    \end{oldappendices}%
}\makeatother

\usepackage{titlesec} 
\titleformat{\section}[block]{\bfseries\sffamily\large}{\thesection. }{0em}{\MakeUppercase} 
\titleformat{\subsection}[block]{\bfseries\sffamily\large}{\thesubsection. }{0em}{} 
\titleformat{\subsubsection}[block]{\large}{}{0em}{\itshape} 

\let\natbibcitet\citet
\renewcommand\citet{\bibpunct{(}{)}{,}{a}{,}{,}\natbibcitet}
\let\natbibcitep\citep
\renewcommand\citep{\bibpunct{(}{)}{;}{a}{,}{;}\natbibcitep}
\newcommand{\bi}{\begin{itemize}}
\newcommand{\ei}{\end{itemize}}
\newcommand{\be}{\begin{equation}}
\newcommand{\ee}{\end{equation}}
\defcitealias{ieo14}{IEO, 2014}

\long\def\symbolfootnote[#1]#2{\begingroup%
\def\thefootnote{\fnsymbol{footnote}}\footnote[#1]{#2}\endgroup}

\makeatletter
\def\ubar#1{\underline{\sbox\tw@{$#1$}\dp\tw@\z@\box\tw@}}
\def\obar#1{\overline{\sbox\tw@{$#1$}\dp\tw@\z@\box\tw@}}
\makeatother

\usepackage{bm}
\usepackage{caption}
\usepackage{subcaption}

\makeatletter\let\p@subfigure\thefigure\makeatother

\floatstyle{plaintop}
\restylefloat{table}

\usepackage{fancyhdr} 
\usepackage{cleveref}
\crefname{chapter}{Chapter}{Chapters}
\crefname{section}{Section}{Sections}
\crefname{subsection}{Section}{Sections}
\crefname{subsubsection}{Section}{Sections}
\crefname{figure}{Figure}{Figures}
\crefname{table}{Table}{Tables}
\crefname{equation}{Equation}{Equations}
\crefname{appendix}{Appendix}{Appendices}
\crefname{appendices}{Appendix}{Appendices}

\crefname{appsec}{Appendix}{Appendices}

\makeatletter

\def\Autoref#1{%
  \begingroup
  \edef\reserved@a{\cpttrimspaces{#1}}%
  \ifcsndefTF{r@#1}{%
    \xaftercsname{\expandafter\testreftype\@fourthoffive}
      {r@\reserved@a}.\\{#1}%
  }{%
    \ref{#1}%
  }%
  \endgroup
}
\def\testreftype#1.#2\\#3{%
  \ifcsndefTF{#1autorefname}{%
    \def\reserved@a##1##2\@nil{%
      \uppercase{\def\ref@name{##1}}%
      \csn@edef{#1autorefname}{\ref@name##2}%
      \autoref{#3}%
    }%
    \reserved@a#1\@nil
  }{%
    \autoref{#3}%
  }%
}
\makeatother
\def\titletext{A Bayesian Gaussian Process Dynamic Factor Model}

\title{\sffamily\huge{\textbf{\titletext}}}
\author{}
\date{}


\def\equationautorefname~#1\null{%
  Eq.~(#1)\null
}
\def\equationautorefname~#1\null{
Eq.~(#1)\null
}

\usepackage[utf8, latin1]{inputenc}                 

\begin{document}

\maketitle
\vspace*{-4.5em} 
\normalsize
\begin{center}
\begin{minipage}{.49\textwidth}
  \large\centering Tony \MakeUppercase{Chernis}\textsuperscript{*}\\[0.25em]
  \footnotesize Bermuda Monetary Authority
\end{minipage}
\begin{minipage}{.49\textwidth}
  \large\centering Niko \MakeUppercase{Hauzenberger}\\[0.25em]
  \footnotesize University of Strathclyde
\end{minipage}

\vspace*{1em}

\begin{minipage}{.49\textwidth}
  \large\centering Haroon \MakeUppercase{Mumtaz}\\[0.25em]
  \footnotesize Queen Mary University of London
\end{minipage}
\begin{minipage}{.49\textwidth}
  \large\centering Michael \MakeUppercase{Pfarrhofer}\\[0.25em]
  \footnotesize WU Vienna
\end{minipage}
\end{center}

\vspace*{1em}

\doublespacing
\begin{center}
\begin{minipage}{\textwidth}

\begin{abstract} \noindent
We propose a dynamic factor model (DFM) where the latent factors are linked to observed variables with unknown and potentially nonlinear functions. The key novelty and source of flexibility of our approach is a nonparametric observation equation, specified via Gaussian Process (GP) priors for each series. Factor dynamics are modeled with a standard vector autoregression (VAR), which facilitates computation and interpretation. We discuss a computationally efficient estimation algorithm and consider two empirical applications. First, we forecast key series from the FRED-QD dataset and show that the model yields improvements in predictive accuracy relative to linear benchmarks. Second, we extract driving factors of global inflation dynamics with the GP-DFM, which allows for capturing international asymmetries.

\medskip
 
\vspace{1ex}

\end{abstract}

\end{minipage}
\end{center}

\begingroup
\renewcommand{\thefootnote}{\fnsymbol{footnote}} 
\footnotetext[1]{The views expressed in this paper are solely those of the authors and may differ from the views of the Bermuda Monetary Authority. No responsibility for them should be attributed to the Bermuda Monetary Authority.}
\endgroup
\singlespacing\vfill\noindent{\footnotesize\textit{Corresponding author}: Haroon Mumtaz (\href{mailto:h.mumtaz@qmul.ac.uk}{h.mumtaz@qmul.ac.uk}). We thank Annika Camehl, Yizhou Kuang, Michele Lenza, and Aubrey Poon, as well as the participants of the \textit{Advances in Macroeconometrics Workshop} in Manchester and the \textit{IAAE 2025} in Turin. Hauzenberger acknowledges funding by the Jubil\"aumsfonds of the OeNB, grant no. 18763.}

\thispagestyle{empty}
\newpage\doublespacing\normalsize\renewcommand{\thepage}{\arabic{page}}
\renewcommand{\footnotelayout}{\setstretch{1.5}}

\section{Introduction}\label{sec:introduction}
Nonlinearities are an important feature of macroeconomic and financial data. In just the last decade, the Global Financial Crisis, COVID-19 pandemic, and central banks reaching their effective lower bound provide examples. Capturing these nonlinearities has become an increasingly important issue---there is widening recognition that they are important for understanding and predicting macroeconomic dynamics.\footnote{For example, a broad strand of literature on macro-financial linkages \citep[e.g.,][]{Brunnermeier2014, ABG2019} or the Phillips curve \citep[e.g.,][]{hamilton2001parametric, BeaudryNBER24, Ball2022inflation, Benigno2023infla} documents instability in the underlying (structural) relationships. This is consequential for effective policy analysis and policy making. Modeling nonlinearities generally has a long tradition in macroeconometrics, with competing approaches ranging from regime-switching \citep[e.g.,][]{hamilton1989new,terasvirta1994specification,sims2006were} to time-varying parameter \citep[TVP, e.g.,][]{primiceri2005time,cogley2005drifts,koop2013large} models.}

In addition to the focus on modeling various forms of nonlinearities, a key feature of modern macroeconometric approaches is their scalability to high-dimensional datasets. These days it is straightforward to download hundreds of macroeconomic time series with a single click; examples for datasets include the US-based FRED-MD/QD \citep[][]{FRED-MD,FRED-QD}, or other well-maintained multi-country databases. Exploiting macroeconomic ``Big Data'' (which typically involves many variables and few observations) to improve structural analysis and forecasting, and, consequently, policymaking, is a focus of researchers and practitioners alike; see \citet{bok2018macroeconomic} for a recent review. And while there is no obvious best way to econometrically model these data \citep[][]{giannone2021economic}, there is a popular front-runner---the dynamic factor model (DFM).

DFMs are a workhorse model of empirical macroeconomics \citep[see,e.g.,][]{aguilar2000bayesian, forni2000generalized, stock2002macroeconomic, kose2003international, GIANNONE2008, ChernisSekkelDFM, kaufmann2019bayesian}.\footnote{\cite{sw2016} and \cite{doz2020dynamic} provide excellent surveys of the earlier literature.} A DFM assumes that there are a few fundamental forces in the economy which explain common dynamics of many time series. These fundamental forces (or latent factors) are usually modeled using a vector autoregression (VAR). A DFM thus compresses the data to work with a more parsimonious VAR, rather than shrinking a large VAR featuring many variables towards a simpler specification, another popular option \citep[see][]{banbura2010large}.

Despite the overall popularity of (linear) DFMs, few papers model any nonlinear relationships between factors and observable variables. As we pointed out above, these may indeed be crucial to understand major macroeconomic fluctuations. Exceptions include DFMs with time-varying loadings \citep{del2008dynamic, mumtaz2012evolving, korobilis2013assessing, zhou2014bayesian}, or DFMs with Markov-switching dynamics \citep{Chauvet1998, chauvet2016dynamic, camacho2018markov}. More recently nonlinear DFMs with a squared/quadratic dynamics in the measurement or state equation \citep{guerron2023financial} have been proposed. However, these studies impose specific functional forms in the context of inferring the latent factors.\footnote{An exception is \cite{Velasco:2024:Chapter3} who uses Bayesian Additive Regression Trees (BART) to model nonlinearities in a DFM. Unlike our approach, she employs a linear approximation to the relationship when estimating the factors. Another related approach is \citet{clark2025nonparametric}, who by contrast use a VAR augmented with static nonlinear factors modeled via BART.} In other words, they impose explicit restrictions on the link between the factors and observed variables.

Machine learning techniques have proven useful for modeling nonlinearities of unknown form in big macroeconomic datasets.\footnote{An incomplete list of examples includes \citet{bassetti2014beta, kalli2018bayesian, farrell2021deep, medeiros2021forecasting, babii2022machine, goulet2022machine, jin2022infinite, huber2023nowcasting, clark2024investigating, chronopoulos2024forecasting, goulet2024macroeconomy, hauzenberger2024bayesian, hauzenberger2024nowcasting, hauzenberger2025gaussian}.} These techniques are appealing for several reasons. First, machine learning approaches are flexible and only require mild assumptions about the form of nonlinearities. Second, these methods are designed to avoid oversimplification, misspecification, and overfitting. Third, these sophisticated methods are well-suited for learning and identifying common patterns in large datasets, enabling efficient information extraction.

This paper introduces a general nonlinear DFM and develops a computationally feasible and fully Bayesian estimation algorithm. As mentioned above, imposing linearity in a typical DFM may be too restrictive, and we thus relax this assumption. Specifically, we propose to use Gaussian processes \citep[GPs,][]{williams2006gaussian}, to obtain a nonparametric Gaussian Process Dynamic Factor Model (GP-DFM). GPs can capture a wide range of possible nonlinear relationships between latent (or observed) factors and high-dimensional data, and they have a successful track record in macroeconomic modeling \citep{clark2024forecasting, hauzenberger2025gaussian}. Compared with other recent nonlinear DFM approaches, our approach is more flexible as this novel framework does not impose any specific type of nonlinearity. Instead, we place a prior (which is compatible with a rich menu of functions, and governed by a distance-based kernel function subject to only a few tuning parameters) directly on the functional relationship between common latent factors and the observed series.

Part of our contribution is bridging the machine learning literature on Gaussian Processes with macroeconometrics in a Big data context (where sample sizes are typically rather small). Specifically, we use GPs in a state space framework \citep{turnerSSM,Frigola2013}. The main novelty of our framework lies in the nonlinear and flexible treatment of the measurement equation, while the state equation is assumed to follow a standard linear VAR. From both a practical and forecasting perspective, the use of a linear VAR in the state equation offers several appealing features.

First, interpretation of the latent common factors is analogous to a standard linear DFM. So, conventional tools from structural and reduced form VAR analysis can be readily used to the linear VAR in the state equation---for example, to produce forecasts as well as impulse response functions (IRFs) for the latent factors quickly. Once the full paths of these forecasts and the IRFs of the factors are known, the nonlinear mapping between observables and latent factors in the measurement equation can be exploited to generate forecasts or IRFs for the observed time series. This treatment results in the ability to calculate structural and reduced form quantities that can be time-varying or state dependent. 

Second, this modeling strategy can also be viewed as a form of nonlinear dimension reduction, where high-dimensional macroeconomic data are assumed to lie on a lower-dimensional manifold or space. Any nonlinearity in the model arises exclusively from the relationship between the latent factors and the observables, which captures the notion of providing a more accurate/precise description of the variation in a large panel of time series. There is a relationship to other dimension reduction techniques such as local linear manifold regression \citep{cheng2013local}, nonlinear principal components \citep[PCs,][]{BAI2008}, deep/nonlinear dynamic factor models and autoencoders, which use neural networks to uncover complex patterns in high-dimensional data \citep{andreini2023deepdynamicfactormodels, hauzenberger2023real, guerron2023financial, klieber2024non, snellman2024nonlinear, luo2025time}. However, unlike existing deep dynamic factor model approaches and popular two-stage procedures, our GP-DFM offers a fully consistent modeling framework. At its core, it represents a Bayesian state space model in which the functional relationship between low-dimensional latent factors and high-dimensional observables is explicitly modeled, with parameters and latent processes with precisely defined priors and posteriors. These posteriors are jointly estimated using a Markov Chain Monte Carlo (MCMC) algorithm (which is part of the contribution of this paper). This framework ensures both model consistency and proper Bayesian uncertainty quantification for all parameters and latent processes in the model, the latter being important for obtaining accurate predictive densities \citep[see][]{geweke2010comparing}.

A fully nonlinear measurement equation is challenging to estimate, as it requires computationally intensive particle filtering techniques. To facilitate Bayesian estimation and provide a computationally feasible option for inference, we use the algorithm proposed in \cite{svensson2016computationally}. This framework involves a linear approximation of the GP using basis functions derived from the spectral densities of the GP kernels \citep{solin2020hilbert}. In addition, we rely on Particle Gibbs with Ancestor Sampling (PGAS) inspired by \citet{lindsten14a}, which greatly improves the efficiency of our algorithm. We further improve computational scalability by proposing an alternative parameterization of the GP kernel that uses an additive structure instead of a multiplicative one. This approach sacrifices some flexibility because it restricts interactions between the factors in the measurement equations, but there are significant computational advantages to this specification. Finally, since our approach uses a fully Bayesian MCMC algorithm, we can straightforwardly implement key model features such as stochastic volatility and shrinkage priors for the VAR process governing the state dynamics.

We provide two different empirical applications. In our first application, we forecast key macroeconomic variables from the FRED-QD dataset. The GP-DFM outperforms linear specifications of the DFM---a workhorse model at many policy institutions. Some of the superior performance is due to the GP-DFM forecasting real activity variables well during the COVID-19 period and the Federal funds rate when it is close to the effective lower bound. Interestingly, we find that with the nonlinear GP-DFM, a smaller number of factors is sufficient to extract the high-dimensional information than when imposing linearity. This supports the notion that in large macroeconomic datasets nonlinear factors can capture richer dynamics than their linear counterparts, which require more components to achieve comparable accuracy. In our second application, we measure international inflation dynamics with GP-DFM, and study nonlinearities and asymmetries arising from shocks to the global component of inflation in a large cross-section of economies.

The rest of the paper is structured as follows. Section \ref{sec:model} lays out the econometric framework. This includes the DFM, GP priors subject to computationally favorable kernel specifications, and the posterior and predictive sampling algorithm. Section \ref{sec:forecastUS} applies the GP-DFM in a forecast horserace with US data, while in Section \ref{sec:inflation} we use the model to capture nonlinearities in international inflation dynamics. Section \ref{sec:conclusions} concludes.

\section{Econometric framework}\label{sec:model}

\subsection{A Gaussian process dynamic factor model (GP-DFM)}\label{ssec:GPDFM}
A general nonlinear DFM relates $N$ observed macroeconomic time series (measurements), $\bm{y}_{t} = (y_{1t},\hdots,y_{Nt})'$, to a set of $D$ common latent factors $\bm f_t = (f_{1t}, \dots, f_{Dt})'$: 
\begin{equation}\label{eq:obseq}
\bm{y}_t = \bm{G}(\bm{f}_t) + \bm{v}_t, \quad \bm{v}_t \sim \mathcal{N}(\bm{0}_N,\bm{R}).
\end{equation}
Equation (\ref{eq:obseq}) is a measurement equation, where an unknown function $\bm G (\bullet):\mathbb{R}^D\rightarrow \mathbb{R}^N$ maps latent inputs $\bm f_t$ to observed outputs $\bm y_t$. In the following, we assume $\bm G(\bm{f}_t) = (g_1(\bm{f}_t), \hdots, g_N(\bm{f}_t))'$ collects equation-specific functions, with $g_i(\bullet): \mathbb{R}^D \rightarrow \mathbb{R}$, for $i = 1, \dots, N$. The idiosyncratic component $\bm v_t$ is Gaussian distributed with zero mean and variance-covariance matrix $\bm R = \text{diag}(r_{1},\dots,r_{N})$. While $\bm{G}(\bm{f}_t)$ captures variation common to all time series in $\bm y_t$, $\bm v_t$ is specific to each time series. Moreover, flexible equation-specific functions allow the common factors $\bm f_t$ to affect each variable differently, adequately reflecting the idea that different time series may feature different types of nonlinearities, captured by possibly different functional forms of $g_i$.

The nonlinear measurement equation is combined with a linear state equation for the factors, which follow a joint VAR($P$) process:\footnote{Note that more general multivariate processes, e.g., $\bm{f}_t = \bm{H}(\bm{f}_{t-1},\hdots,\bm{f}_{t-p}) + \bm{\varepsilon}_{t}$, can and have been used as state equations in nonlinear state space models \citep[e.g.,][]{Frigola2013,guerron2023financial}; see \citet{marcellino2024bookchapter} for a recent review of the macroeconometric literature using such models with (observed) macro-data. For the reasons we outlined in the Introduction (i.e., striking a balance between flexbility, computational efficiency, general ease of use and interpretability), we impose that $\bm{H}(\bullet)$ is linear a priori and leave nonlinear extensions for future research.}
\begin{equation}
 \bm{f}_t = \sum_{p=1}^{P} \bm{A}_{p} \bm{f}_{t-p} + \bm{\varepsilon}_{t}, \quad \bm{\varepsilon}_{t} \sim \mathcal{N}(\bm 0_D, \bm{Q}), \label{eq:state-factor}
\end{equation}
with coefficient matrices $\bm A = (\bm A_1, \dots, \bm A_P)$ that relate $\bm f_t$ to their $P$ lags. In addition, $\bm{\varepsilon}_t = (\varepsilon_{1t}, \dots, \varepsilon_{Dt})'$ are zero mean state innovations with variance-covariance matrix $\bm{Q}$.\footnote{In the empirical application, we also investigate whether allowing for stochastic volatility (SV), with a time-varying $\bm Q_t$ rather than a constant $\bm Q$, improves forecast accuracy \citep[see][]{C2011JBES, CR2015JAE}.} 

In terms of features of this general framework, a few considerations are noteworthy. Eqs. (\ref{eq:obseq}) and (\ref{eq:state-factor}) define a state space model. Specifically, $\bm f_t$ can be viewed as a lower dimensional representation of the high-dimensional data $\bm y_t$, which strikes a balance between explaining most of the variation in the data and parsimony. In this regard, a critical assumption is $D \ll N$. 

A standard DFM assumes a linear mapping, $\bm{G}(\bm{f}_t) = \bm{\Lambda} \bm{f}_t$, with $\bm{\Lambda}$ being an $N \times D$ loadings matrix. In such a case, Eqs. (\ref{eq:obseq}) and (\ref{eq:state-factor}) resemble a Gaussian linear state space model and for estimation standard algorithms such as forward sampling backward smoothing \citep[FFBS,][]{carter1994gibbs, fruhwirth1994data} or the precision sampler \citep{chan2009efficient} can be used. However, this linearity assumption may be too restrictive and is an assumption we wish to relax. To do so, one might consider incorporating nonlinear features, e.g., of the form $\phi(\bm f_t) = (\bm f_t', (\bm f_t \odot \bm f_t)')'$; a related approach is discussed in \citet{guerron2023financial}.

Related to this so-called ``feature space'' \citep[see][chapter 2]{williams2006gaussian} two points are worth highlighting. 
First, $\phi(\bm f_t) = (\bm f_t', (\bm f_t \odot \bm f_t)')'$, or other simple transformations, would denote a deterministic projection of $\bm f_t$ onto $\phi(\bm f_t)$. However, we use a general form for $\phi(\bm f_t)$ which, by virtue of being more flexible, can approximate deterministic transforms but does not require knowing the appropriate transform ahead of time. Second, recall $\bm f_t$ is latent. Hence, any $\phi(\bm f_t)$ that includes nonlinear transforms of $\bm f_t$ leads to a nonlinear state space model, for which standard sampling algorithms can no longer be used. In short, our approach does not require pre-specifying the form of nonlinearity and we provide an estimation algorithm which is feasible for nonlinear measurement equations. The remaining sub-sections of the econometric framework will focus on these two considerations and carefully outline the contribution of this paper.

\subsection{Gaussian processes (GPs) as the measurement equation}
The key innovation of our model is the measurement equation features unknown nonlinearities. We choose GPs since they have a proven track record in macroeconomic (and many other) applications, and because of several properties that will result in a computationally feasible estimation algorithm. The key feature of GPs is that they model the covariance between observations through a simple kernel function. Kernel functions usually only have few parameters, but they are very expressive. Despite their sparse parametrization, GPs can indeed approximate a wide variety of functions.\footnote{There are theoretical equivalencies between neural nets and GPs \citep{Neal1996, lee2018deepneuralnetworksgaussian}.} 

We begin describing the model, starting with the $i^\text{th}$ measurement equation: 
\begin{equation*}
 y_{it} = g_{i}(\bm{f}_{t})+v_{it}, \quad v_{it}\sim \mathcal{N}(0,r_{i}).
\end{equation*}
We assume a GP prior on each observable-specific function: 
\begin{equation}\label{eq:GPprior}
g_{i}(\bm{f}_t) \sim \mathcal{GP}(0,\mathcal{K}_{i}(\bm{f}_t,\bm{f}_t)),
\end{equation}
where $\mathcal{K}_{i}(\bullet)$ is a suitable covariance function (called the kernel) which we define below. Eq. (\ref{eq:GPprior}) denotes a prior over all possible functions $g_{i}$ that fit the $i^\text{th}$ equation well, without explicitly taking a stance of the functional form of $g_i$ and thus the feature space of $\bm f_t$. Through an appropriate choice of kernel we can allow for an infinite set of basis functions and feature space $\phi(\bm f_t)$. This means a GP can approximate any arbitrary continuous function, see also \citet{williams2006gaussian}. 

This paper focuses on two distinct versions of the commonly used squared exponential kernel. One variant is multiplicative across covariates and the other additive:
\begin{subequations} \label{eq:sqexpK}
\begin{align}
    \mathcal{K}_{i,\KMul}(\bm{f}_t,\bm{f}_\tau) &= \xi_i \cdot \prod_{j = 1}^D \exp\left(-\frac{1}{2} \frac{(f_{jt} - f_{j\tau})^2}{\ell_{ij}^2}\right)\label{eq:sqexpK_M} \\
    \mathcal{K}_{i,\KAdd}(\bm{f}_t,\bm{f}_\tau)  &= \xi_i \cdot \sum_{j = 1}^D \exp\left(-\frac{1}{2} \frac{(f_{jt} - f_{j\tau})^2}{\ell_{ij}^2}\right), \label{eq:sqexpK_A}
\end{align}
\end{subequations}
with hyperparameters $\bm{\theta}_{i} = (\xi_i, \{\ell_{ij}\}_{j = 1}^D)'$ defining covariances between two periods $t$ and $\tau$. Here, 
$\xi_i$ denotes an unconditional variance parameter and $\{\ell_{ij}\}_{j = 1}^D$ refer to factor-specific length scales. We label the former variant in Eq. (\ref{eq:sqexpK_M}) the {multiplicative} (\texttt{multi}) GP-DFM, and the latter in Eq. (\ref{eq:sqexpK_A}) the {additive} (\texttt{hybrid}) GP-DFM. Both variants define a stationary covariance function appropriate for macroeconomic data. However, they differ in the interactions between inputs (factors). The additive kernel sums the squared exponential terms instead of multiplying them. This difference has implications for computation and modeling flexibility, which we discuss in more detail below. 

In the following, we first discuss the implications of GPs with a generic stationary kernel in the measurement Eq. (\ref{eq:obseq}). If $\bm f_t$ were known, this would denote an otherwise standard GP regression, with inference being straightforward. However, $\bm f_t$ is latent which substantially complicates estimation and inference.  To see this, define $\bm{g}_i = (g_{i}(\bm{f}_1),\hdots,g_{i}(\bm{f}_T))'$, then the GP takes the form:
\begin{equation*}
    \bm{g}_i \sim \mathcal{N}(\bm{0}_T,\mathcal{K}_i(\bm{F},\bm{F}')),
\end{equation*}
with $\mathcal{K}_i(\bm{f}_t,\bm{f}_\tau)$ being the $(t,\tau)^\text{th}$ entry in $\mathcal{K}_i(\bm{F},\bm{F}')$. Considering the weight-space view of our GP model in the $i^{\text{th}}$ equation, we obtain:
\begin{equation*}
\bm y_i = \bm W_i \bm \eta_i + \bm v_i, \quad \bm \eta_i \sim \mathcal{N}(\bm 0, \bm I_T).
\end{equation*}
Here, $\bm W_i$ denotes the lower Cholesky factor of $\mathcal{K}_i(\bm{F},\bm{F}') = \bm W_i \bm W_i'$, with $w_{i, t \tau}$ referring to the $(t, \tau)^\text{th}$ element of $\bm W_i$. Note that $\mathcal{K}_i(\bm{F},\bm{F}')$ denotes a full (symmetric) $T \times T$ variance-covariance matrix, implying that the measurement equation relates $y_{it}$ to the full history of latent states $\bm f_t$:
\begin{equation}\label{eq:Wview}
y_{it} = \sum_{\tau = 1}^{t} w_{i, t\tau} \eta_{i\tau} + v_{it}.
\end{equation}
As alluded to by the weight space view, the measurement equation is rather involved, rendering a fully Bayesian approach feasible but computationally cumbersome \citep[see, e.g.,][]{Frigola2013}.


\subsection{Reduced-rank approximation of the GP}\label{subsec:apprxGP} 
In our paper, we approximate the GP using a set of nonlinear basis functions. These basis functions are based on the spectral density of the kernels in Eq. (\ref{eq:sqexpK}), see \citet{svensson2016computationally, solin2020hilbert, riutort2023practical}. Loosely speaking, this facilitates estimation by breaking the dependence on the full history of latent factors $\{\bm f_\tau\}_{\tau = 1}^{t}$, relying instead only on transformations of $\bm f_t$. In general, this approximation is enabled by the fact that any stationary covariance function can be represented in terms of their spectral densities $\mathcal{S}_i(\bm{\omega})$. Using eigenfunction expansion, any stationary covariance function with inputs $\bm{f}_t$ can then be expressed as:
\begin{equation}
\mathcal{K}_i(\bm{f}_t,\bm{f}_\tau) = \sum_{m=1}^{\infty} \mathcal{S}_i(\sqrt{\bm{\lambda}_m}) \phi_m(\bm{f}_t)\phi_m(\bm{f}_\tau),  \label{eq:GPspec}
\end{equation}
with $\mathcal{S}_i$ denoting the spectral density, $\bm{\lambda}_m$ are vectors of eigenvalues and $\phi_m(\bm{f}_t)$ are eigenfunctions of the Laplacian operator that is used to establish the approximation in domain $\Omega$. More formally, $\Omega \in [-L_1,L_1] \times \hdots \times [-L_D,L_D]$ (i.e., $\Omega \subset \mathbb{R}^D$) describes the support on which the GP approximations are valid \citep[see also][for details]{solin2020hilbert}. Crucially, these eigenvalues and eigenfunctions do not depend on the specific choice of the kernel or its hyperparameters---only the spectral density $\mathcal{S}_i$ does.

Following \citet{riutort2023practical}, the kernels can then be approximated using a set of $M$ basis functions:\footnote{The covariance function within $\Omega$ with inputs $\bm{f}_t,\bm{f}_\tau \in \Omega$ may be written as $\mathcal{K}_{i}(\bm{f}_t,\bm{f}_\tau) = \sum_{m=1}^{\infty} \mathcal{S}_{i}(\sqrt{\bm{\lambda}_m}) \phi_m(\bm{f}_t)\phi_m(\bm{f}_\tau)$. We may truncate this sum so that it can be used to obtain the weight space representation of the GP in Eq. (\ref{eq:GPapprox}).}
\begin{equation}
    g_i(\bm{f}_t) \approx \sum_{m=1}^{M} \phi_m(\bm{f}_t)c_{im}, \quad c_{im}\sim\mathcal{N}\left(0,\mathcal{S}_{i}(\sqrt{\bm{\lambda}_m})\right).\label{eq:GPapprox}
\end{equation}
These approximations, for $i = 1, \dots, N$, allow us to rewrite Eq. (\ref{eq:obseq}) as:
\begin{equation}\label{eq:obsapprox}
\bm{y}_t = \bm{C}\bm{\Phi}(\bm{f}_t) + \bm{v}_t,
\end{equation}
with the $N\times M$-matrix $\bm{C}$ comprising the weights $c_{im}$ across equations $i = 1,\hdots,N,$ and basis functions $m = 1,\hdots,M,$ and the basis functions are stored in an $M \times 1$-vector $\bm{\Phi}(\bm{f}_t) = (\phi_1(\bm{f}_t),\hdots,\phi_{M}(\bm{f}_t))'$. Note that the derivations above are applicable to any generic stationary kernel. Next, we focus on distinct aspects that arise for the multiplicative and additive variants that we mentioned earlier. Indeed, the two squared exponential covariance functions in Eq. (\ref{eq:sqexpK}) are both stationary and therefore the associated spectral densities are Fourier duals:
\begin{equation*}
\mathcal{S}_{i}(\bm{\omega}) =
\begin{cases}
\xi_i (2\pi)^{D/2} \left(\prod_{j = 1}^D \ell_{ij}\right) \cdot \exp\left(-\frac{1}{2} \sum_{j=1}^D \ell_{ij}^2 \omega_j^2\right), &\text{for } \mathcal{K}_{i,\KMul},\\
\xi_i (2D^2\pi)^{1/2} \left(\sum_{j = 1}^D \ell_{ij} \cdot \exp\left(-\frac{1}{2} \ell_{ij}^2 \omega_j^2\right)\right), &\text{for } \mathcal{K}_{i,\KAdd},
\end{cases}
\end{equation*}
with a vector $\bm{\omega} = (\omega_1,\hdots,\omega_D)'$ in the frequency domain.

\inlinehead{Multiplicative squared exponential kernel} Let $\mathbb{S}\in\mathbb{R}^{M \times D}$ be the matrix of all possible combinations of univariate eigenfunctions across dimensions (i.e., all $D$-tuples), and we obtain for $m = 1,\hdots,M$:
\begin{align*}
\bm{\lambda}_m = \left(\left\{\frac{(\pi\mathbb{S}_{mj})^2}{4L^2}\right\}_{j=1}^D\right)', \quad \phi_m(\bm{f}_t) = \prod_{j=1}^D L^{-1/2} \sin\left(\lambda_{\mathbb{S}_{mj}}^{1/2} (f_{jt} + L)\right).
\end{align*}
With a $D$-dimensional input space, the total number of eigenvalues and eigenfunctions used for the approximation is equal to the number of possible combinations of univariate eigenfunctions across all dimensions, i.e., $M = \tilde{M}^D$ where $\tilde{M}$ is the number of basis functions for each dimension. In case of the multiplicative squared exponential kernel, the approximation changes the computational complexity to $\mathcal{O}((T + 1) M^D)$ from $\mathcal{O}(T^3)$, a significant saving as long as the number of factors remains low. 

\inlinehead{Additive squared exponential kernel}
Large macroeconomic datasets often require a moderate number of factors to fit the data. For example, \cite{FRED-QD} find seven or eight factors in the full FRED-QD dataset, which would be rather computationally demanding when used with the multiplicative squared exponential kernel. For this reason, we introduce introduce the additive squared exponential kernel as a more computationally efficient alternative.

Start by noting that the GP with this kernel is $g_{i}(\bm{f}_t) \sim \mathcal{GP}(0,\sum_{j = 1}^D \mathcal{K}_{i}(f_{jt},f_{jt}))$, which can be split into the sum of $D$ univariate GPs,
$y_{it} = g_{i}(f_{1t}) + \dots + g_{i}(f_{Dt}) + v_{it}$. Let $\tilde{M}$ now denote the number of basis functions for each dimension $(j = 1, \dots, D)$. Then the total number of eigenvalues/eigenfunctions used for the approximation is equal to $M = \tilde{M}^D$ for the multiplicative kernel, while it is $M = D \times \tilde{M}$ for the additive kernel. This is because we can consider $\mathbb{S}\in\mathbb{R}^{\tilde{M}}$ for each dimension separately and simply sum over the factor-specific basis functions given the additive structure: 
\begin{align*}
\bm{\lambda}_{mj} = \left(\left\{\frac{(\pi\mathbb{S}_{mj})^2}{4L^2}\right\}_{j=1}^D\right)', \quad \phi_m(\bm{f}_t) = \sum_{j=1}^D L^{-1/2} \sin\left(\lambda_{\mathbb{S}_{mj}}^{1/2} (f_{jt} + L)\right).
\end{align*}

This seems to be a classic trade-off between model flexibility and computational feasibility. The main limitations of the additive kernel variant is that it ignores interactions between latent factors, potentially underfitting highly complex relationships. On the other hand, it comes with the benefit of substantially improved scalability in the number of latent factors ($D$). One of the empirical contributions of our paper is investigating this trade-off in macroeconomic data.

\subsection{Prior setup}\label{subsec:prior} 
As we adopt a fully Bayesian approach, we need to assign prior distributions to all unknown parameters in the model. For the measurement equation, we closely follow recent developments in the Gaussian process literature \citep[see][]{vandvaart2008bayesian, bhattacharya2014anisotropic, lindsten14a, hauzenberger2025gaussian}. Specifically, for the equation-specific (homoskedastic) measurement errors in Eq. (\ref{eq:obseq}), we assume independent inverse Gamma priors, $r_{i} \sim \mathcal{G}^{-1}\left(\nu_r, S_r\right)$, with $\nu_r = 3$ and $S_r = 0.3$ for all $i = 1, \dots, N$ equations. For the hyperparameters of the Gaussian processes kernels, we use Gamma priors on the unconditional variances, $\xi_i \sim \mathcal{G}\left(\nu_\xi, S_\xi \right)$, and Gamma priors on the inverse length-scale parameters, $\ell^{-2}_{ij} \sim \mathcal{G}\left(\nu_\ell, S_\ell\right)$. In the empirical application, We set $\nu_\xi = \nu_\ell = 0.5$ and $S_\xi = S_\ell = 0.5$, which is a weakly informative choice.

To specify priors for the parameters in the state equation, Eq. (\ref{eq:state-factor}), we follow the recent macroeconomic DFM/VAR literature and adopt global--local shrinkage priors where appropriate \citep[see, e.g.,][]{huber2019adaptive, kaufmann2019bayesian, fruhwirth2025sparse}. Specifically, we use horseshoe \citep[HS,][]{carvalho2010horseshoe} priors, as they have been shown to work well in diverse contexts virtually free of tuning. Any hyperparameters associated with these priors are updated in a data-driven manner, making the approach suitable for VARs of any size and providing a highly adaptive prior specification in the state equation. We defer technical details on the prior setup for the state equation to Section \ref{app:prior} of the Appendix.
   
\subsection{A Particle Gibbs sampling algorithm}\label{subsec:Gibbs}
To sample from the joint posterior distribution of our nonlinear state space model we rely on a Particle Gibbs approach. This section sketches the main steps of the algorithm, and additional details are provided in Section \ref{app:MCMC} of the Appendix. Our Markov Chain Monte Carlo (MCMC) involves three main steps/blocks:

\begin{enumerate}[leftmargin =*]
\item \textbf{Sample common latent factors.} We use Particle Gibbs with Ancestor Sampling (PGAS) to update $\{\bm f_t\}_{t = 1}^T$ from its conditional posterior distribution \citep{Andrieu2010, lindsten14a}. \cite{Andrieu2010} show how a version conditioning on a fixed trajectory for one of the particles can be used to produce draws that result in a Markov kernel with a target distribution that is invariant. However, the usual problem of path degeneracy in particle filters can result in poor mixing in the original version of this approach (especially for earlier periods in the sample). Recent development suggest that small modifications of this baseline algorithm can largely alleviate this problem. 

In particular, \cite{lindsten14a} propose the addition of a step that involves sampling the ``ancestors" or indices associated with the particle that is being conditioned on. They show that this results in a substantial improvement in the mixing of the algorithm even with only a few particles. As explained in \cite{lindsten14a}, ancestor sampling breaks the reference path into segments allowing the particle system to collapse onto a new higher probability path. In the absence of ancestor sampling the particle system tends to collapse to the reference path; PGAS is outlined in detail in the first sampling block of Section \ref{app:MCMC} in the Appendix. 

\item \textbf{Sample unknown parameters in the measurement equation.}
Most parameters can be sampled on an equation-by-equation basis (i.e., independently over $i = 1, \dots, N$) using well-known conditional posterior distributions. The exception are the GP hyperparameters, which require equation-specific Metropolis-Hastings (MH) steps. We update:
\begin{enumerate}[leftmargin =*, label=(\roman*)]
    \item \textbf{Variance of the idiosyncratic components $r_{i}$}: Using a conditionally conjugate inverse Gamma prior results in an inverse Gamma conditional posterior.
    \item \textbf{Coefficients in the observation equation $\bm c_i = (c_{i1}, \dots, c_{iM})'$}: As shown in \citet{svensson2016computationally}, the reduced form approximation outlined in Section \ref{subsec:apprxGP} implies a specific form for the conditional posterior of the coefficients in the observation equation. The conditional posterior is a multivariate Gaussian.
    
    \item \textbf{Kernel parameters $\bm \theta_{i}$}: To draw $\bm \theta_{i}$, we use an MH step with a random walk proposal. This amounts to proposing a candidate from suitable proposal, and computing the usual acceptance probably.
\end{enumerate}
Specifics about these conditional posterior distributions and any associated moments are provided in the second block of Section \ref{app:MCMC} in the Appendix.

\item \textbf{Sample unknown parameters in the state equation.} Conditional on a full history of $\bm f_t$, we sample the parameter in the state equation using the algorithm proposed in \cite{carriero2022corrigendum}. These sampling steps for the state equation are outlined in detail in the third block of Section \ref{app:MCMC} in the Appendix. 
\end{enumerate}
Typically none of the individual parameters of multivariate time series models are of primary interest, but functions of them are. In terms of computing forecasts and IRFs with the GP-DFM model, it suffices to note that the procedure is relatively straightforward. For the latent factors, we can iteratively project them $\tau$ steps ahead, $\bm f_{t+1}, \dots, \bm f_{t+\tau}$, using conventional tools for the linear VAR in the state equation (\ref{eq:state-factor}). Once the path of the latent factors is known, we can use the approximated GP measurement equation (\ref{eq:obsapprox}) to obtain forecasts/IRFs for the endogenous variables.

\section{Forecasting the US Macroeconomy}\label{sec:forecastUS}
In a first application, we assess the overall performance of the GP-DFM in an extensive forecasting exercise using data for the US macroeconomy. We use the popular FRED-QD dataset \citep[see][]{FRED-QD} which is a large information set spanning a broad array of quarterly macroeconomic data. Our evaluation sample is long enough to include several US macroeconomic events of importance: the dot-com bubble, the great financial crisis (GFC), and the COVID pandemic. We benchmark our main model specifications against a standard linear DFM which is a workhorse model in policy institutions for forecasting \citep{GIANNONE2008, ChernisSekkelDFM}. Section \ref{ssec:FCSTsetup} provides details on the forecasting setup as well as the competing specifications and Section \ref{ssec:FCSTresults} summarizes the main forecasting results.

\subsection{Forecasting setup and competing specifications}\label{ssec:FCSTsetup}
The FRED-QD dataset covers 248 quarterly time series of which we select $N = 105$ to replicate the information content of \cite{stock2012disentangling}. This rich  macroeconomic dataset covers real activity, price, and financial variables. All variables are transformed to stationarity as suggested in \cite{FRED-QD}. For a list of variables used and transformations applied, see Appendix \ref{app:data}. We define real output (\texttt{GDPC1}), non-farm employment (\texttt{PAYEMS}), consumer prices (\texttt{CPIAUCSL}) and the Federal funds rate (\texttt{FEDFUNDS}) as target variables for the forecasting exercise. We use an expanding window estimation scheme. The estimation sample starts in 1965Q1 and is recursively updated each quarter over the hold-out sample which spans 1992Q1 to 2023Q4. We consider forecast horizons of one-quarter, one-year and two-years ahead. For our target variables, we assess both joint and marginal density forecast accuracy. We use the energy score (ES) to evaluate the joint predictive densities and  the continuous rank probability score (CRPS) to evaluate the marginal predictive densities \citep{GR2007JASA}. 

In the forecasting exercise, we consider 16 different model specifications, which vary across key modeling choices. For all models we use $P = 4$ lags in the state equation. The margins we explore are linearity versus nonlinearity (or kernel choice) of the DFM: \texttt{linear DFM} for the linear specification, \texttt{GPDFM-A} for the GP-DFM with an additive kernel, and \texttt{GPDFM-M} for the GP-DFM with a multiplicative kernel. Moreover, we vary the number of factors ($D = \{2, 4, 8\}$) and consider either homoskedastic variances or SV of the state innovations. The out-of-sample forecasting exercise is used to assess all of these choices because it is robust and we do not have clear theoretical guidance for them. Finally, we need to decide on a few GP-DFM specific choices. For the multiplicative kernel, we set $\tilde{M} = 8$ for $D = 2$ and $\tilde{M} = 4$ for $D = 4$, and we do not consider $D = 8$.\footnote{For the multiplicative kernel, $D = 8$ would result in a total of $M = \tilde{M}^8$ basis functions. Thus, setting $\tilde{M} = 4$ yields $M = 65536$, which is computationally burdensome.} For the additive kernel, we set $\tilde{M} = 8$ for all values of $D$. The boundary condition $L$ is specified in a semi-automatic manner as $1.2$ times the maximum absolute value of the first $D$ PCs.\footnote{We have empirically tested different $\tilde{M}$ values and our choices seem to be sensible ones in terms of the accuracy of the GP approximation.}



\subsection{Out-of-sample forecasting evidence}\label{ssec:FCSTresults}

\inlinehead{Summary of findings} Before zooming into the details we provide a broad overview of the forecasting results. There are four main take aways. First, overall, the nonlinear DFM specifications consistently perform well and outperform the linear DFM, providing some improvements over the linear benchmark when considering both joint and marginal forecast losses. Second, the computationally efficient additive kernel method performs similarly to the more flexible multiplicative kernel, indicating that there is little trade-off between computation speed and forecast accuracy. Third, using only a relatively small number of factors is sufficient to forecast our target variables accurately. Forth, as is common in the macroeconomic literature, allowing for SV improves forecast accuracy. 

\begin{figure}[h!]	
\caption{Joint forecast performance.\label{fig:es}} 
\centering
\begin{minipage}{\textwidth}
\centering
(a) Full evaluation sample from 1992Q1 to 2023Q4
\end{minipage}
\begin{minipage}{\textwidth}
\centering
\includegraphics[width=0.7\paperwidth]{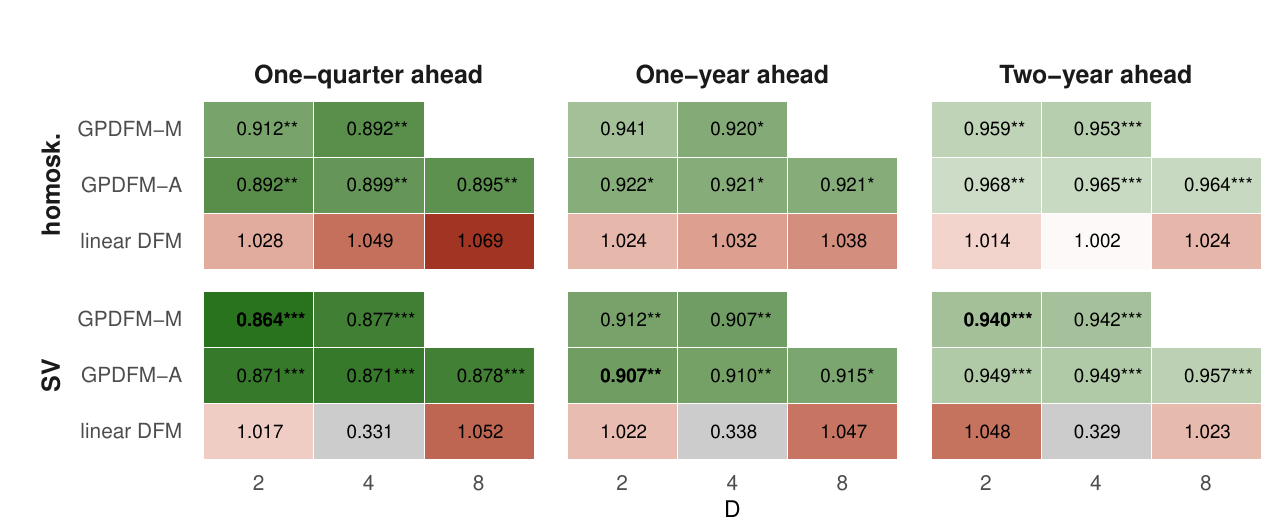}
\end{minipage}
\begin{minipage}{\textwidth}
\centering
(b) Pre-COVID evaluation sample from 1992Q1 to 2019Q4
\end{minipage}
\begin{minipage}{\textwidth}
\centering
\includegraphics[width=0.7\paperwidth]{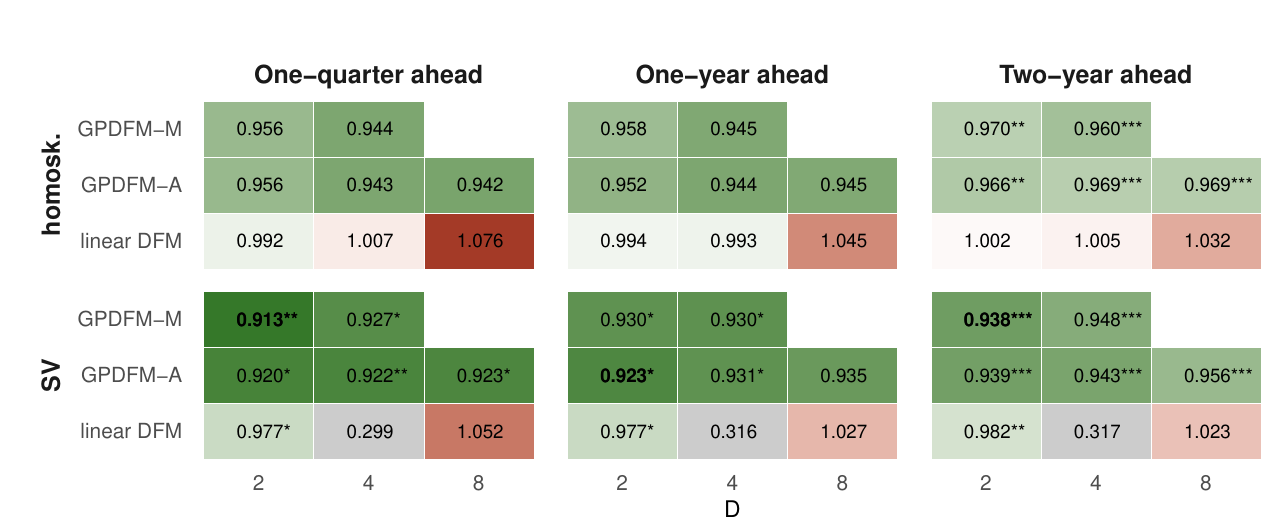}
\end{minipage}
\begin{minipage}{\textwidth}
\centering
(c) Pre-GFC evaluation sample from 1992Q1 to 2007Q4
\end{minipage}
\begin{minipage}{\textwidth}
\centering
\includegraphics[width=0.7\paperwidth]{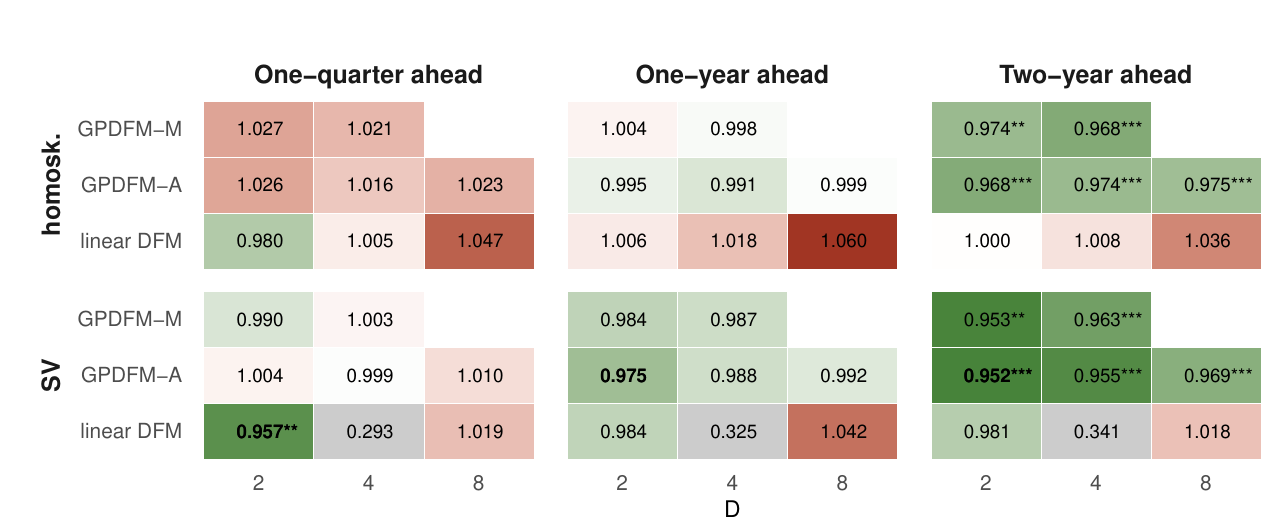}
\end{minipage}
\caption*{\scriptsize \textbf{Notes:} Energy scores \citep[ES,][]{GR2007JASA} are used as a measure of joint forecast performance across all target variables. The gray-shaded entries refer to the absolute ES scores of a linear DFM with $D = 4$ and SV, which serves as the benchmark. A ratio smaller than one implies that the respective model outperforms the benchmark (indicated by a green-shaded cell), while a ratio greater than one implies that the respective model is outperformed by the benchmark (indicated by a red-shaded cell). Asterisks indicate statistical significance of forecast accuracy gains relative to the benchmark at the $1\%$ ($^{***}$), $5\%$ ($^{**}$), and $10\%$ ($^{*}$) significance levels, using a one-sided \cite{DM1995JBES} test. The best performing specification for each horizon is in bold.}
\end{figure}

\begin{figure}[h!]	
\caption{Marginal one-quarter ahead forecast performance.\label{fig:crps1}} 
\begin{minipage}{\textwidth}
\centering
(a) Full evaluation sample from 1992Q1 to 2023Q4
\end{minipage}
\begin{minipage}{\textwidth}
\centering
\includegraphics[width=0.7\paperwidth]{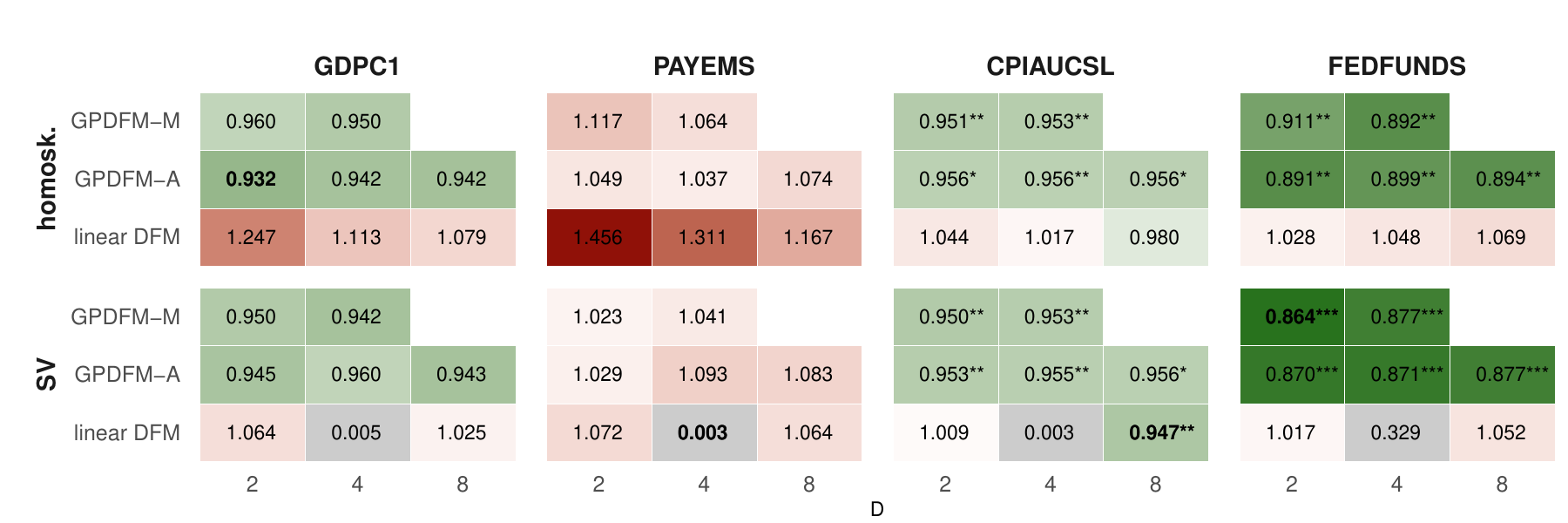}
\end{minipage}
\begin{minipage}{\textwidth}
\centering
(b) Pre-GFC evaluation sample from 1992Q1 to 2007Q4
\end{minipage}
\begin{minipage}{\textwidth}
\centering
\includegraphics[width=0.7\paperwidth]{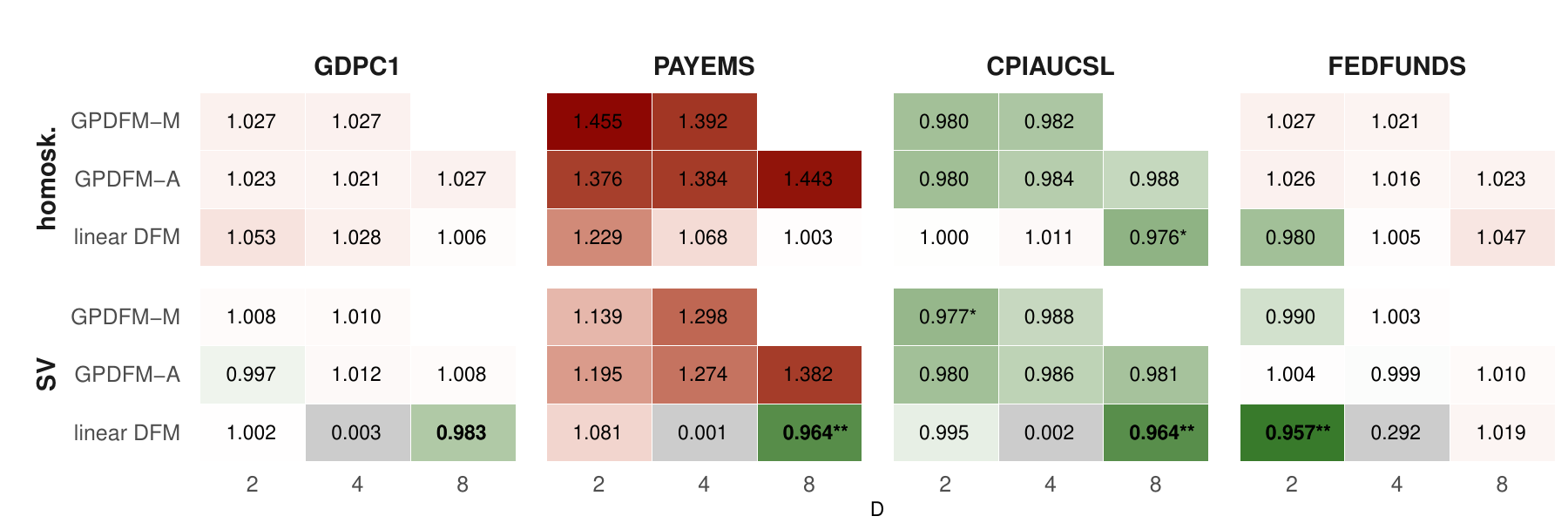}
\end{minipage}
\caption*{\scriptsize \textbf{Notes:} Continuous rank probability score \citep[CRPS,][]{GR2007JASA} are used as a measure of marginal forecast performance for each target variable. The gray-shaded entries refer to the absolute CRPS scores of a linear DFM with $D = 4$ and SV, which serves as the benchmark. A ratio smaller than one implies that the respective model outperforms the benchmark (indicated by a green-shaded cell), while a ratio greater than one implies that the respective model is outperformed by the benchmark (indicated by a red-shaded cell). Asterisks indicate statistical significance of forecast accuracy gains relative to the benchmark at the $1\%$ ($^{***}$), $5\%$ ($^{**}$), and $10\%$ ($^{*}$) significance levels, using a one-sided \cite{DM1995JBES} test. The best performing specification for each horizon is in bold.}
\end{figure}

\begin{figure}[h!]	
\caption{Marginal higher-order forecast performance (full evaluation sample).\label{fig:crps2}} 
\centering
\begin{minipage}{\textwidth}
\centering
(a) One-year ahead 
\end{minipage}
\begin{minipage}{\textwidth}
\centering
\includegraphics[width=0.7\paperwidth]{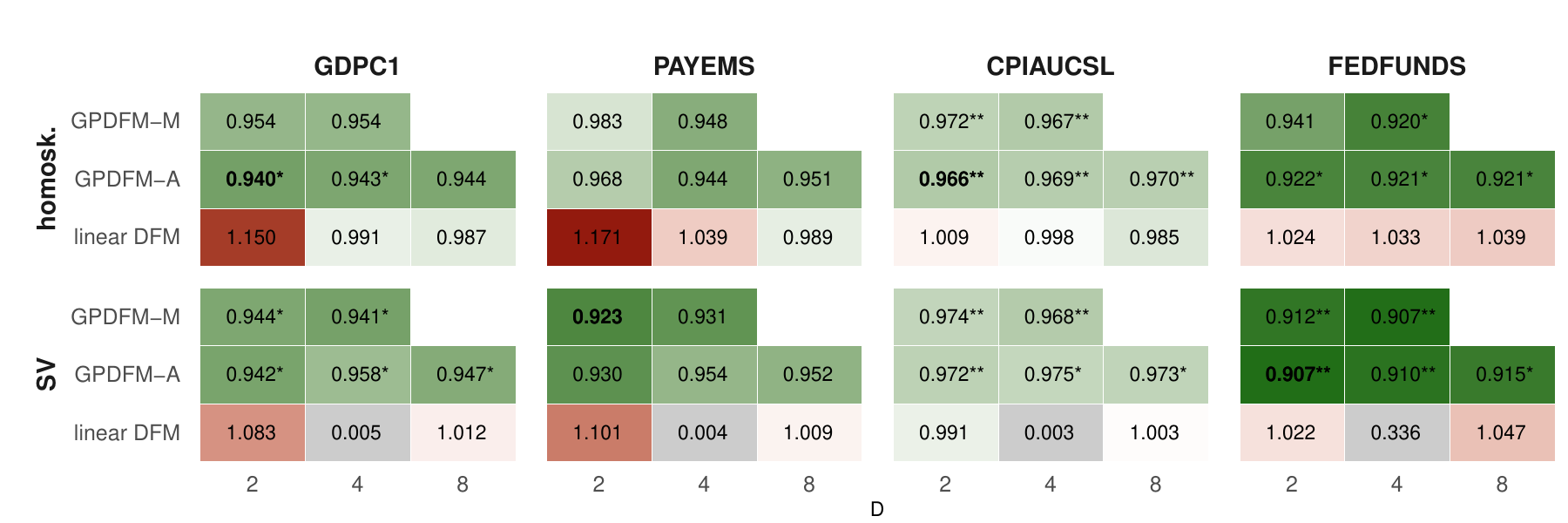}
\end{minipage}
\begin{minipage}{\textwidth}
\centering
(b) Two-year ahead
\end{minipage}
\begin{minipage}{\textwidth}
\centering
\includegraphics[width=0.7\paperwidth]{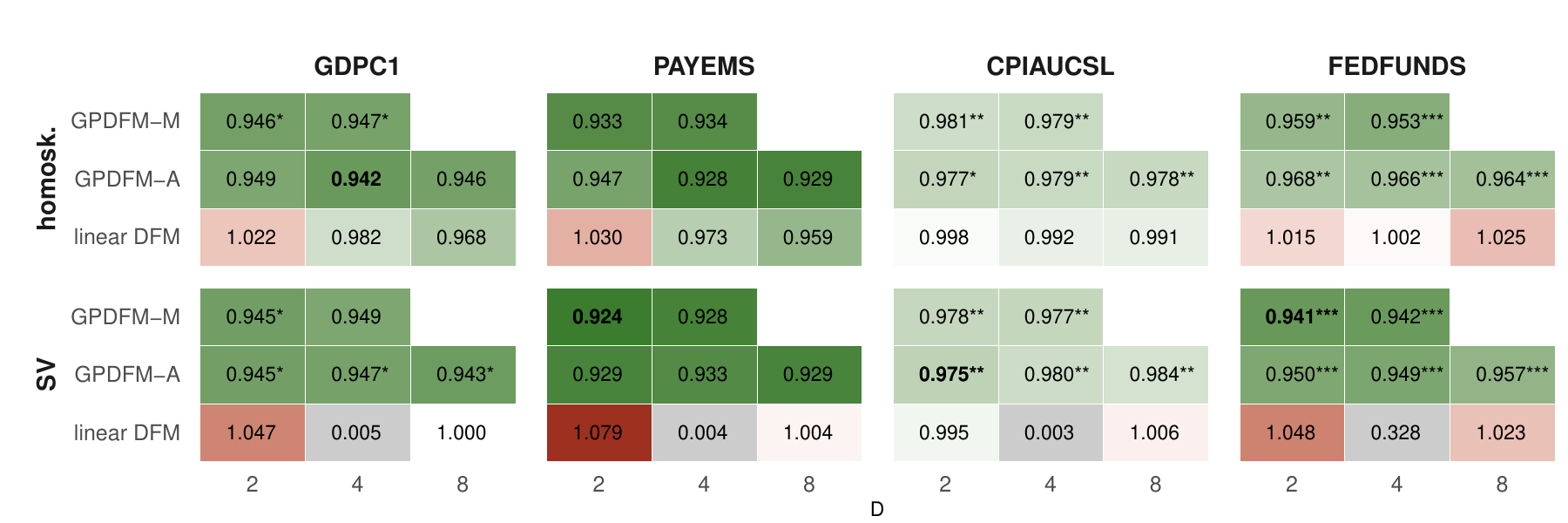}
\end{minipage}
\caption*{\scriptsize \textbf{Notes:} Continuous rank probability score \citep[CRPS,][]{GR2007JASA} are used as a measure of marginal forecast performance for each target variable. The gray-shaded entries refer to the absolute CRPS scores of a linear DFM with $D = 4$ and SV, which serves as the benchmark. A ratio smaller than one implies that the respective model outperforms the benchmark (indicated by a green-shaded cell), while a ratio greater than one implies that the respective model is outperformed by the benchmark (indicated by a red-shaded cell). Asterisks indicate statistical significance of forecast accuracy gains relative to the benchmark at the $1\%$ ($^{***}$), $5\%$ ($^{**}$), and $10\%$ ($^{*}$) significance levels, using a one-sided \cite{DM1995JBES} test. The best performing specification for each horizon is in bold.}
\end{figure}

In the following, Figs. \ref{fig:es} to \ref{fig:crps2} in this section report forecast evaluation metrics for both the joint forecast performance and the marginal forecast performances. For ease of readability, all figures follow a common convention. The benchmark model is a linear DFM with $D = 4$ and SV, and its absolute forecast score is reported directly. For other model specifications, we show the ratio relative to the benchmark. Ratios below one indicate superior performance relative to the benchmark (indicated by green-shaded cells), while ratios above one indicate inferior performance (indicated by red-shaded cells). Statistical significance is assessed using a one-sided \cite{DM1995JBES} test, with asterisks denoting the significance of gains relative to the benchmark at the $1\%$ ($^{***}$), $5\%$ ($^{**}$), and $10\%$ ($^{*}$) level. For each horizon, the best performing specification is highlighted in bold. 


\inlinehead{Joint forecast performance over different evaluation periods} Figure \ref{fig:es} shows the joint forecast performance across horizons for three different evaluation samples. In panel (a) we consider the full evaluation sample, in panel (b) we stop prior to the COVID period in $2019$Q$4$, and in panel (c) we stop prior to the GFC in $2007$Q$4$. This strategy should help identify forecast gains over time. Recall our target variables are real output (\texttt{GDPC1}), non-farm payroll employment (\texttt{PAYEMS}), consumer prices (\texttt{CPIAUCSL}) and the Federal funds rate (\texttt{FEDFUNDS}); the losses in this context are based on the joint predictive distribution of these variables.

We first focus on panel (a) in Fig. \ref{fig:es}, which shows the joint forecast performance for the full evaluation sample and serves as the main measure of model performance. As suggested in the key summarized findings mentioned above, we find that a GP-DFM variant with two factors and SV in the state equation consistently outperforms all other specifications across horizons. This model variant exhibits significant forecast improvements around $6$ to $14\%$ lower losses (as indicated by the dark green shaded cells).

When we vary the evaluation sample, we observe significant differences in the relative forecast performance of the GP-DFMs. Comparing panels (a) and (b) of Fig. \ref{fig:es} shows that including the COVID period systematically lowers ES ratios for our GP variants relative to the benchmark (but also relative to other linear competitors). This suggests that the impressive performance is partly driven by the period of the pandemic, and indicates that a flexible, nonlinear GP-DFM improves forecast accuracy over a more restrictive linear DFM during this exceptional period. Moreover, the improvements of the GP-DFM variants are largest at shorter horizons, outperforming linear competitors by larger margins. The best performing model improves upon the benchmark by nearly $14\%$ for the full evaluation sample (and around $9\%$ for the pre-COVID sample) at the one-quarter ahead horizon, while the improvement for the two-year ahead horizon is about $6\%$, irrespective of whether the evaluation stops before COVID. 

Comparing panels (b) and (c) of Fig. \ref{fig:es}, it is evident that the gains arise at least partially in the context of the GFC and the zero lower bound (ZLB). Indeed, the shorter (pre-GFC) period can be considered a relatively calmer economic period, and while there are some improvements for GP-DFM variants for the one-quarter and one-year horizon, the linear DFM benchmark is only occasionally outperformed, and typically not significantly so. For the two-year ahead horizon, however, the nonlinear model is more accurate, with significantly lower losses.

\inlinehead{Marginal forecast performance} Figures \ref{fig:crps1} and \ref{fig:crps2} show the marginal forecast performance for the four target variables for different subsamples and across several forecast horizons. For one-quarter ahead forecasts we specifically look at the performance pre-GFC and the full sample. We do so because particularly at this horizon, GP-DFM shows sizable differences in joint relative performance, and the marginal perspective allows to investigate the sources of this accuracy premium also in terms of the cross-section of variables. For higher-order forecast horizons, we report the marginal scores only for the full evaluation sample in the interest of saving space.

Focusing first on the results at the one-quarter ahead horizon in Fig. \ref{fig:crps1}, the CRPS ratios echo joint forecast performance patterns, and show that accuracy varies substantially across evaluation samples. Models with SV and a low number of factors perform particularly strongly for the full sample. All GP-DFM variants significantly outperform the linear benchmark at a $1\%$ significance level (as well as the other linear competitors) for the Federal funds rate (\texttt{FEDFUNDS}). Here, the best performing specification, a GP-DFM with two factors, a multiplicative kernel, and SV, outperforms the benchmark by almost $14\%$. This is likely responsible for a sizable portion of the joint forecast performance. But this is not the sole source of the larger forecast accuracy gains for the full evaluation sample, as the GP-DFM also forecasts real output and inflation well. For real output (\texttt{GDPC1}), the best performing model is a GP-DFM equipped with two factors, an additive kernel, and homoskedastic errors, improving about $7\%$ upon the benchmark (but these improvements are not statistically significant). For inflation (\texttt{CPIAUCSL}), most GP-DFM variants significantly outperform the benchmark at the $5\%$ significance level, showing forecast performance similar to the best performing specification (in this case, a linear DFM with eight factors and SV). By contrast, for payroll employment (\texttt{PAYEMS}), there are no improvements.

Turning to the pre-GFC evaluation sample in panel (b), a linear DFM equipped with SV appears highly competitive and is difficult to outperform using GPs during this period (which largely coincides with the comparatively calm Great Moderation). Indeed, for the pre-GFC evaluation sample, virtually all GP-DFM variants perform similarly as the linear benchmark for \texttt{GDPC1}, \texttt{CPIACUSL}, and \texttt{FEDFUNDS}. However, the proposed nonlinear models struggle with forecasting \texttt{PAYEMS} at short horizons, even more so in ``normal'' economic times.

Moving to the one-year and two-year ahead horizons in Fig. \ref{fig:crps2}, we see that a GP-DFM variant is consistently the best performing model across all variables. Specifically, the GP-DFM with SV and two factors performs very well. While we observe strong performance for the Federal funds rate similar to the short horizon, forecast performance for real activity variables generally improves at the longer horizons. In addition, inflation forecasts are better by a small but statistically significant amount (at the $5\%$ significance level). The comparatively poor performance for short horizon forecasts of payroll employment also vanishes, with GP-DFM exhibiting more accurate forecasts than the benchmark (although these are not statistically significantly better).

\inlinehead{Properties of real output growth forecast densities} In the final part of this section, we showcase the features of the predictive densities of a GP-DFM variant using real output growth around the COVID pandemic as a means to detect outliers and nonlinearities. We can highlight what drives differences in forecast performance by comparing the predictive densities of the benchmark model against a GP-DFM variant.

\begin{figure}[!htbp]		
\caption{Assessment of one-year ahead real output growth predictive densities.\label{fig:dens_diff}}
\centering
\includegraphics[width=0.8\textwidth]{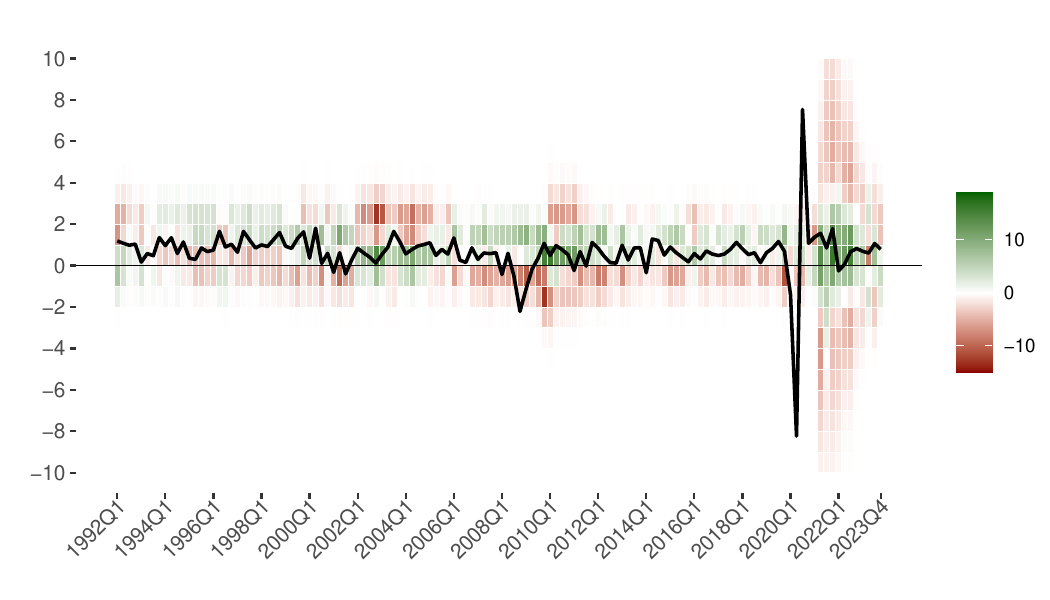}
\caption*{\scriptsize \textbf{Notes:} This figure shows the difference in predictive probabilities between the GP-DFM variant with four factors, an additive kernel, and SV, and the linear DFM with four factors and SV \citep[see][]{DSZ2022JE}. The grid ranges from $-10\%$ to $10\%$ in increments of $1\%$. Green (red) shaded cells indicate that the GP-DFM increases (reduces) probability relative to the linear DFM in a certain region. The black solid line denotes the realized real output growth rate (\texttt{GDPC1}).}
\end{figure}

Figure \ref{fig:dens_diff} shows the difference between the predictive densities of a GP-DFM variant with four factors, an additive kernel, and SV versus the linear DFM with four factors and SV.\footnote{Such a comparison between predictive densities has been proposed in \cite{DSZ2022JE}.} To ensure a fair comparison, we focus on the same number of factors and allow both models to feature SV. By inspecting this figure, we can see that the GP-DFM predictive densities place more mass around the realized outcomes, while the linear DFM is more dispersed. This cannot be attributed to SV, which features in both models. Instead, the GP-DFM's tighter forecast densities are likely a key factor behind the improved forecast performance. Specifically, this is evident during the calmer post-COVID period (from 2021 to 2023) following the sharp downturn and rebound in output growth at the onset of the pandemic.

\section{Decomposing the drivers of global inflation}\label{sec:inflation}
Our second application demonstrates how to use the model for a semi-structural analysis, by investigating the importance of nonlinearities in macroeconomic data. Specifically, we explore the co-movement in cross-country consumer price index (CPI) inflation rates in the spirit of \cite{mumtaz2012evolving} and \cite{Kose2025}. We estimate the following model:
\begin{equation}
\pi_{it} = g_{i}(f_{\Wf, t}) + s_i \cdot h_{i}(f_{\Df, t}) + (1 - s_i) \cdot h_{i}(f_{\Ef, t}) + e_{it}. \label{dfm_1}
\end{equation}
The panel of (quarterly) CPI inflation rates, $\pi_{it}$, is taken from the World Bank inflation database compiled by \cite{Kose2025}, which runs from $1970$Q$1$ to $2023$Q$4$ and includes $N_{\Df} = 27$ developed countries as well as $N_{\Ef} = 34$ emerging economies (EMDEs), such that $i=1,2,\hdots,N_{\texttt{D}},N_{\texttt{D}}+1,\hdots, N,$ with $N = N_{\Df} + N_{\Ef}$. Let $\mathbb{I}(\bullet)$ refer to an indicator function, and define $s_i = \mathbb{I}(i \leq N_{\Df})$. That is, $s_i = 1$ if the respective country is among the developed countries and $0$ when it is an emerging economy. Following \cite{Kose2025}, inflation is decomposed into a global ``world'' factor (indicated with $\Wf$) that is common to all countries ($f_{\Wf, t}$) and regional factors that are specific to country groups, namely developed countries ($f_{\Df, t}$) and EMDEs ($f_{\Ef, t}$). As each inflation series might show a certain of idiosyncratic persistence, we assume that the idiosyncratic errors in the observation equation follow an AR($2$) process: $e_{it}= \sum_{q=1}^{2} \rho_{iq}e_{it-q}+ v_{it}$, $v_{it} \sim N(0,r_{i})$, which is a straightforward extension.

Unlike the previous literature, we allow the relationship between factors and observables to be nonlinear via the unknown functions $g_i(\bullet)$ and $h_i(\bullet)$. Our measurement equation allows for state-dependence and time-varying contributions from factors to observables. Note Eq. (\ref{dfm_1}) corresponds to using an additive kernel similar to the previous section, with $g_{i}(\bullet)$ specific to $f_{\Wf, t}$ and $h_{i}(\bullet)$ specific to the relevant regional factor. We assume a GP with a squared exponential kernel for each function that are approximated using $\tilde{M} = 8$ basis functions. The regional factors are disentangled by imposing exclusion restrictions as in Eq. (\ref{dfm_1}). That is, the regional factor for developed economies does not affect EMDEs and vice versa. We normalize the sign and scale of the factor by using a tight prior on the first row of loadings in each regional group.\footnote{The prior is obtained using an initial estimate of the weights obtained conditional on principal component estimate of the factors.} The latent factors $\bm f_{t}= (f_{\Wf, t}, f_{\Df, t}, f_{\Ef, t})'$ are assumed to evolve according to a VAR(4) process, see Eq. (\ref{eq:state-factor}), with $P = 4$.  

\subsection{Features of the GP-DFM inflation factors}
\begin{figure}[!htbp]
\caption{Estimated inflation factors of the GP-DFM.\label{factors}}
\centering
\begin{minipage}{0.32\textwidth}
\centering
\includegraphics[width=1\textwidth]{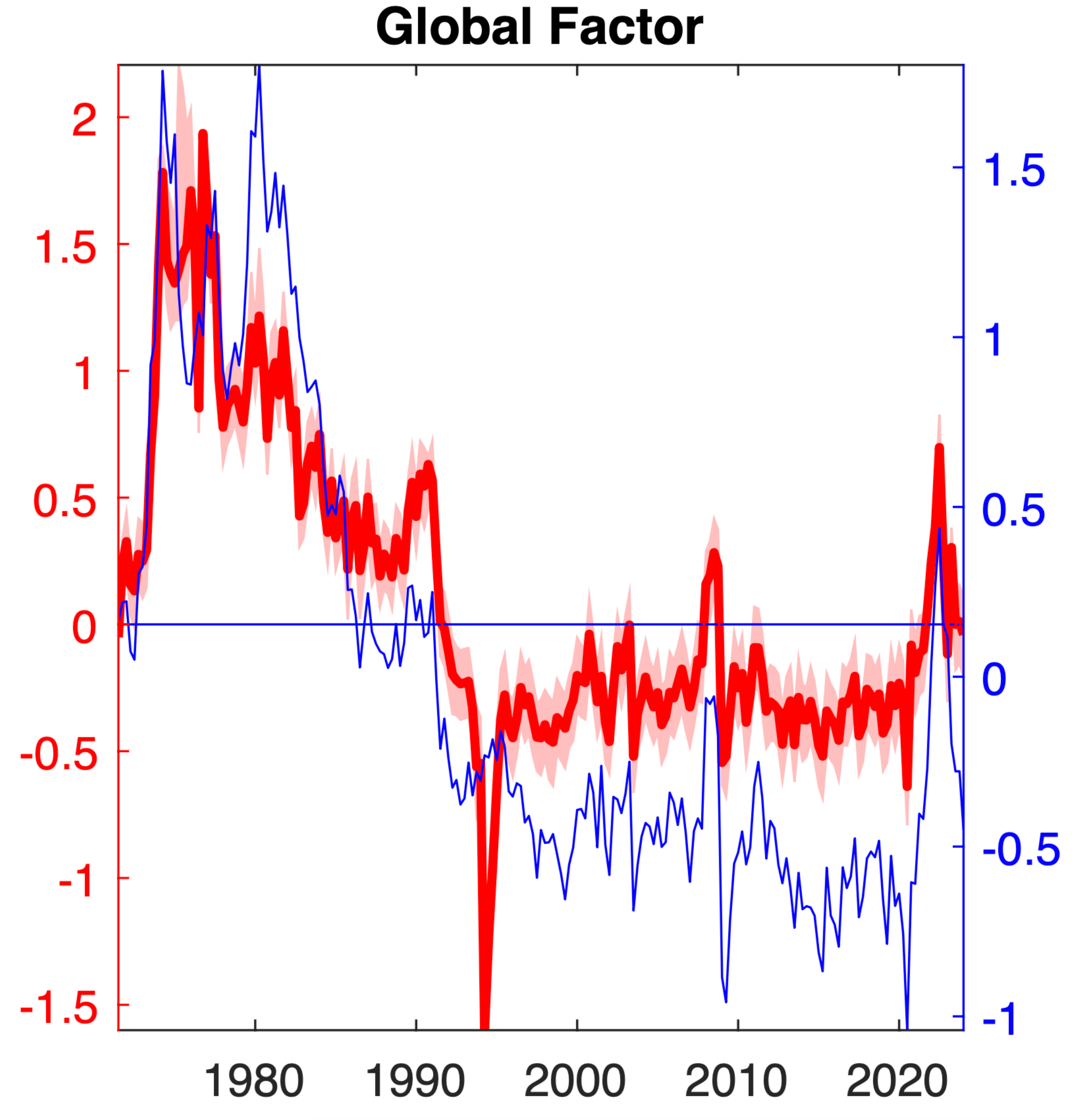}
\end{minipage}
\begin{minipage}{0.32\textwidth}
\centering
\includegraphics[width=0.98\textwidth]{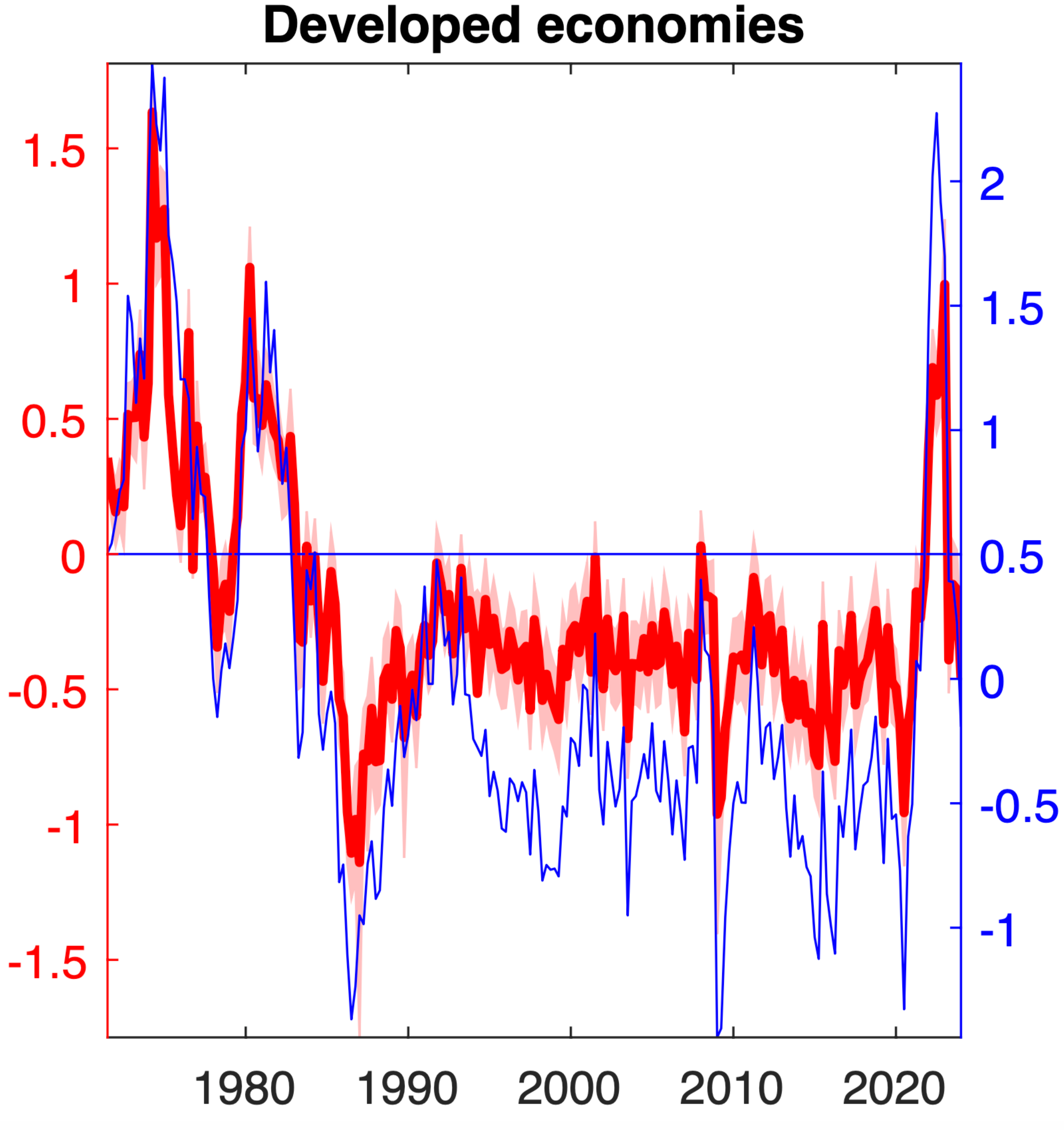}
\end{minipage}
\begin{minipage}{0.32\textwidth}
\centering
\includegraphics[width=1.02\textwidth]{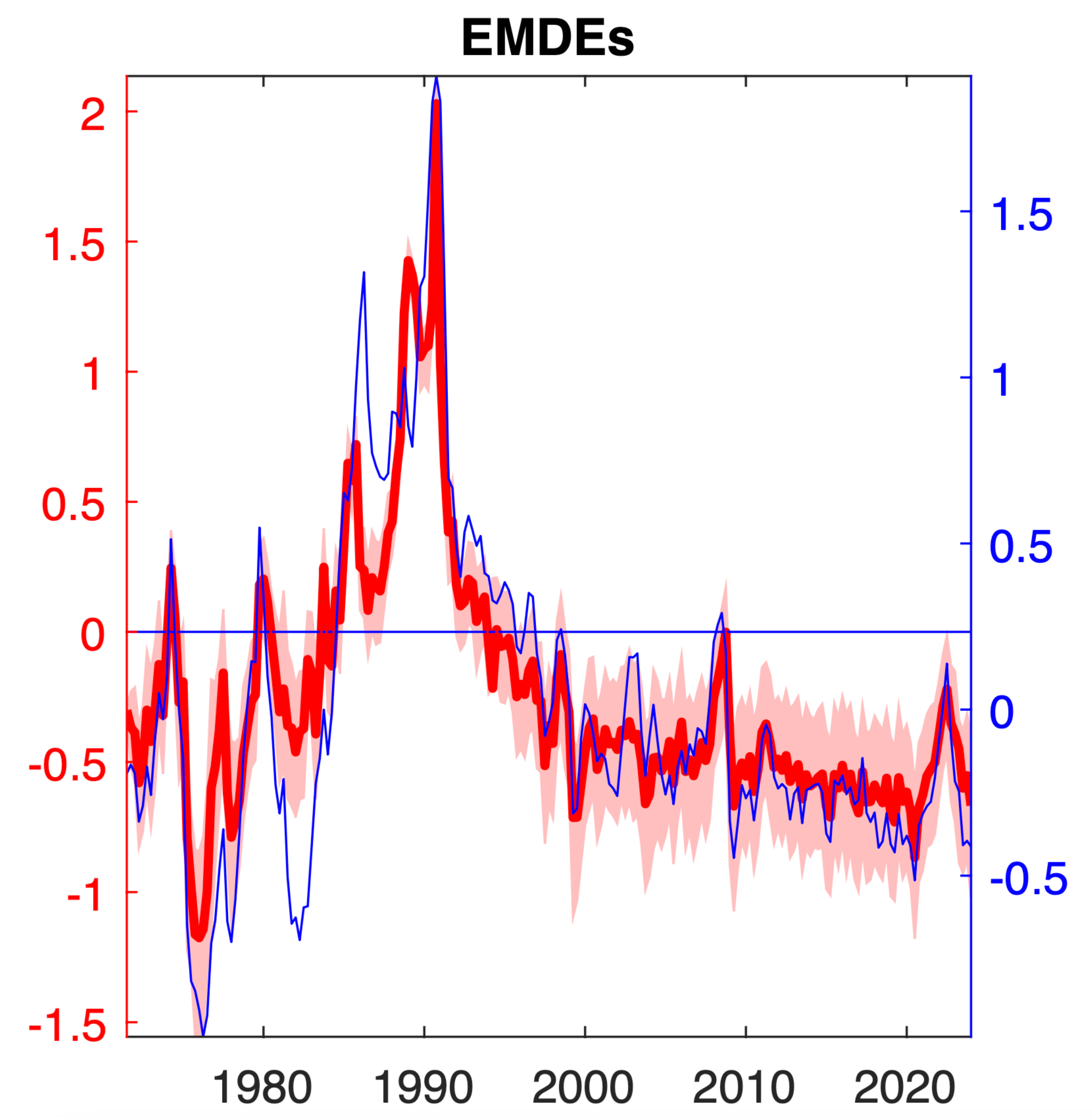}
\end{minipage}
\caption*{\scriptsize \textbf{Note:} The figure displays the posterior estimates of the factors from the GP-DFM model. The red lines display the median estimates while the red shaded areas indicate the $68\%$ confidence intervals. The blue lines represent the posterior medians of the factors from a linear DFM.}
\end{figure}

Figure \ref{factors} plots the estimated posterior distribution of the three factors. The global factor has a distinct peaks in the earlier part of the sample in 1974Q1 correspond to the aftermath of the oil shock. After 1990, the global factor displays a declining trend consistent with the Great Moderation. This came to an end with a sharp rise in $2022$Q$1$, possibly associated with supply disruptions in the post-COVID period.

The factor specific to developing economies is persistently high during the mid-1970s and then during the early 1980s. The decline in this factor in 1985 precedes the fall in global factor and is much sharper. In the subsequent period, inflation in developing countries appears to have been stable with the abrupt disruption after 2021. The variation in inflation specific to EMDEs appears radically different. The main peaks in the EMDE factor occurs in 1989 and 1990 capturing the aftermath of debt crises and hyperinflation in several countries.

It is interesting to the note that while the estimated factors from the benchmark model are broadly similar to that obtained from the linear DFM, there are noticeable differences. For example, the global factor from the benchmark model has a smaller peak in 1980 when compared to the estimate from the linear DFM and displays sharper movements in 1994 and 2008. The developed country factor in the benchmark case displays less variation than the linear factor during the 1990s and 2000s while the EMDE factor has a smaller peak during the mid-1980s than the linear counterpart.

\subsection{The importance of nonlinearities}
This section investigates if there are meaningful nonlinearities between factors and observables. We study this by varying the size and sign of shock applied to the factors and compute generalized forecast error variance decompositions (GFEVD). A linear model would show no difference in the GFEVD for different size/sign shocks whereas if nonlinearities are important there would be substantial differences.  For example, a large negative global inflation shock would have a proportionally different impact than a small positive shock in a nonlinear model.

More specifically, we compute generalized impulse responses \citep[GIRFs,][]{koop1996impulse}, using a Cholesky decomposition of the reduced form covariance matrix to orthogonalize the structural disturbances \citep[see also][for related discussions]{pfarrhofer2025scenario}. In the state equation VAR, we order the global factor first, followed by the developed economy and the EMDE factor, respectively. The GIRFs are calculated using the estimated factors in every fourth quarter of the sample as initial conditions. We use the generalised forecast error variance decomposition (GFEVD) to estimate the contribution of the global and group specific factors to country-specific inflation rates. The GFEVD is calculated using the method proposed by \cite{lanne2016generalized} whereby the GIRFs are used in the standard formula for the FEVD of a linear VAR model. This ensures that the decomposition adds up to one, while retaining properties such as dependence on initial conditions and on the size and sign of the shock being considered.

\begin{landscape}
\begin{figure}
\centering
\caption{Forecast error variance decomposition of country-specific inflation series.\label{fevd}}
\begin{minipage}{0.48\linewidth}	
\centering 
(a) Developed economies
\end{minipage}
\begin{minipage}{0.48\linewidth}	
\centering 
(b) EMDEs
\end{minipage}
\begin{minipage}{0.48 \linewidth}	
\centering 
\includegraphics[width=1\textwidth]{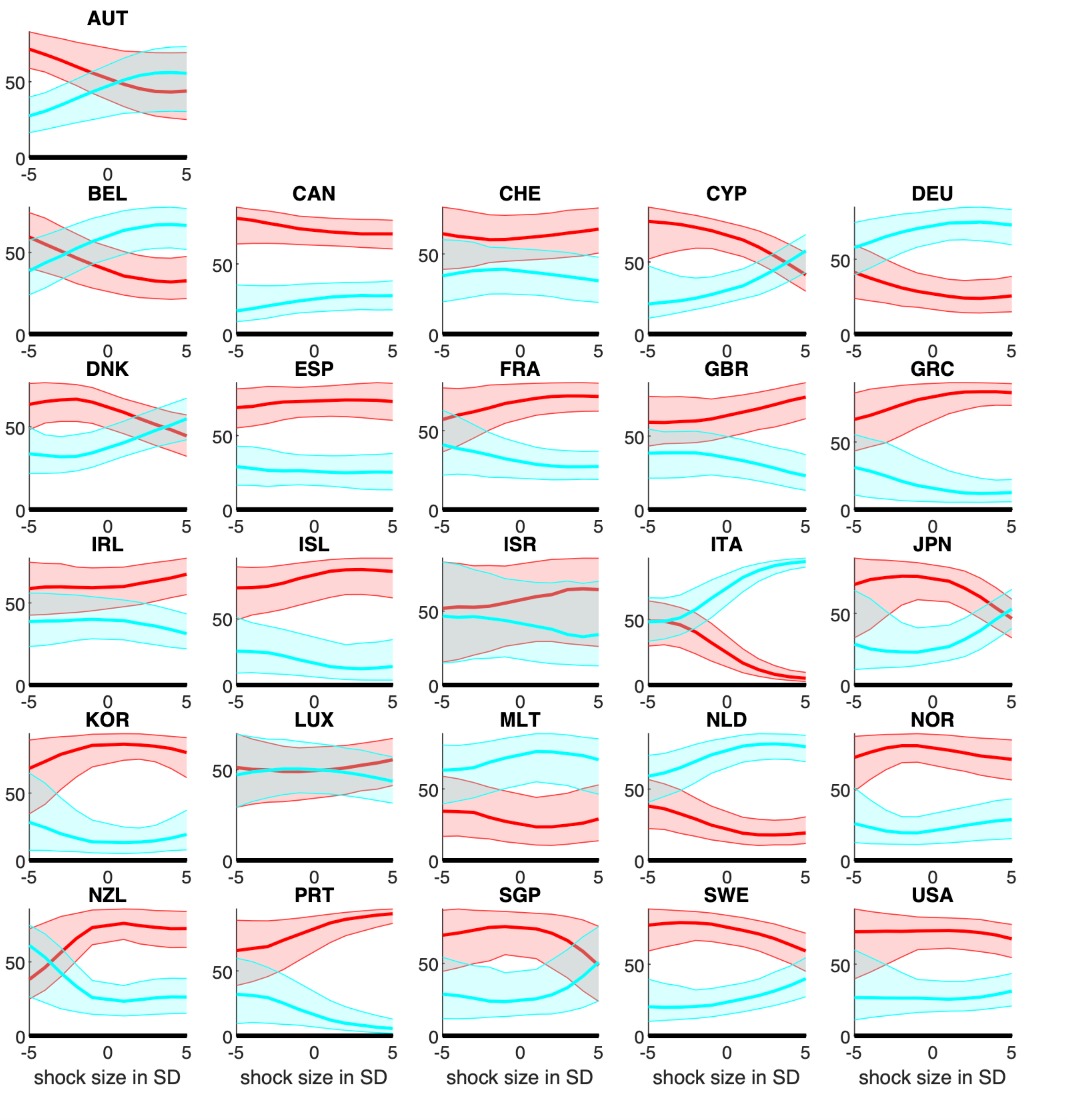}
\end{minipage}
\begin{minipage}{0.48\linewidth}	
\centering 
\includegraphics[width=1\textwidth]{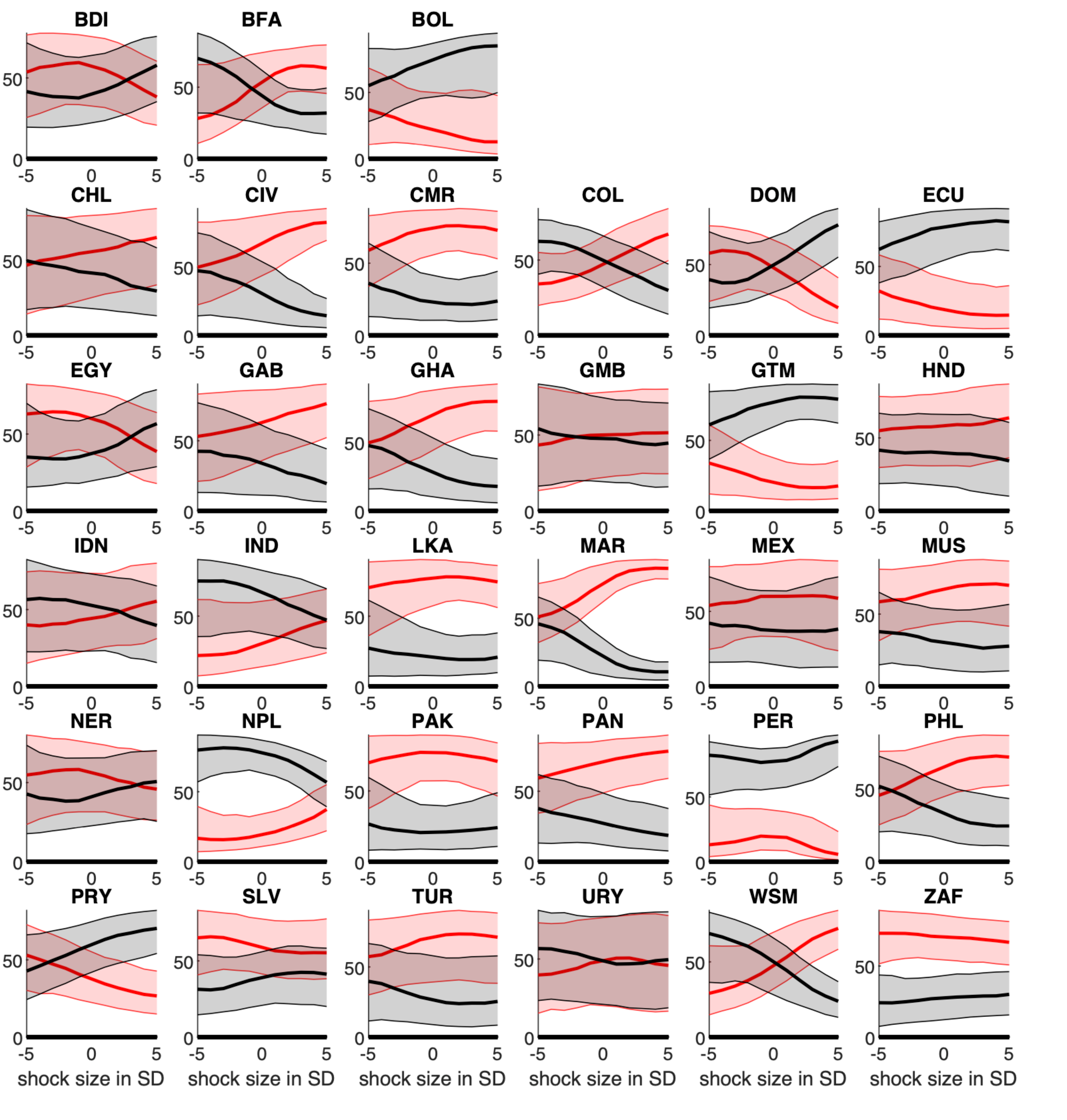}
\end{minipage}
\caption*{\scriptsize \textbf{Note:} The figure displays the contribution to the forecast error variance of inflation at the one-year horizon of shocks to the global factor (red), developed country factor (blue) for different shock sizes. The contributions are averaged across initial conditions. The solid lines are medians while the shaded areas represent the $68\%$ confidence intervals.}
\end{figure}
\end{landscape}

Panel (a) in Fig. \ref{fevd} shows that the contribution of the global and regional factors to the forecast error variance of inflation in developed countries. There is a clear dependence of the contributions on the sign and size of the shock. For example, the contribution of the world (regional) factor is lower (higher) for large positive shocks in Austria, Belgium, Cyprus, Denmark, Italy, Netherlands and Japan. The opposite result appears to hold for countries such as the UK, Greece, New Zealand, and Portugal. 

Panel (b) Figure \ref{fevd} shows that for many EMDEs the importance of the world factor also depends on the size and sign of the shock. This is especially apparent for countries such as Burkina Faso, Colombia, Morocco and Ghana. For these countries, the regional factor contributes more to inflation fluctuations when shocks to this factor are negative.

\section{Conclusions}\label{sec:conclusions}
In this paper we propose a class of nonlinear dynamic factor models using Gaussian Processes. Our first contribution is developing an estimation approach. Furthermore, we show that this model performs well and is easy to use for structural applications. The estimation algorithm we develop makes estimation of the GP-DFM feasible by addressing many of the drawbacks of Gaussian process models. For example, a GP models have difficulties scaling and filtering unobserved states in nonlinear models is difficult. We show how to solve these problems with an efficient estimation approach using Hilbert space approximations and particle sampling with ancestor sampling. Additionally, we show how to further increase computational speed using an additive kernel so that the model can handle moderate numbers of factors. Succinctly, we make using a nonlinear dynamic factor model feasible in macroeconomic applications.

We assess the GP-DFM in a forecasting exercise using US macroeconomic data. We show that the GP-DFM with SV outperforms linear versions of the model. Specifically, it performs well when there are nonlinearities in the data such as at the effective lower bound and during COVID. On average, its performance is driven by its more precise predictive densities. Finally, there is little cost for using an additive kernel, which excludes interactions between factors, suggesting that interaction terms are not an important type of nonlinearity in this dataset. 

We also demonstrate how such a model could be used for a semi-structural analysis. We show that our specification with a nonlinear measurement equation and linear state equation makes for easy inference. The linear state equation allows us to apply structural VAR techniques to the factors while the nonlinear measurement equation allows for flexibility in how the factor maps to observables (state-dependence and time-variation). Our example investigates the co-movement in cross-country CPU inflation rates. First we find differences between the factors estimated in a linear DFM and a GP-DFM. Second, we investigate the importance of nonlinearities by examining how the size and sign of shocks affect contributions to the forecast error variance decomposition. This shows the importance of allowing for nonlinearities.

{\setstretch{0.9}
\addcontentsline{toc}{section}{References}
\bibliographystyle{custom.bst}
\bibliography{lit}}\normalsize\clearpage\doublespacing

\begin{appendices}\crefalias{section}{appsec}
\begin{center}
{\LARGE\sffamily\textbf{Online Appendix:\\\titletext}}
\end{center}

\setcounter{page}{1}
\setcounter{section}{0}
\setcounter{equation}{0}
\setcounter{footnote}{0}

\renewcommand\thesection{\Alph{section}}
\renewcommand\theequation{\Alph{section}.\arabic{equation}}
\renewcommand\thefigure{\Alph{section}.\arabic{figure}}
\renewcommand\theequation{\Alph{section}.\arabic{equation}}

\section{Technical Appendix}\label{app:tech}
The technical appendix provides further details on the priors used in the state equation (\ref{eq:state-factor}) and on well-known steps of the Particle within Gibbs sampling algorithm.

\subsection{Shrinkage prior setup for the state equation}\label{app:prior}
To specify the priors on the parameters the parameters in the state equation, Eq. (\ref{eq:state-factor}), define $\bm A = (\bm A_1, \dots, \bm A_P)$ and decompose $\bm Q = \bm \Psi^{-1} \bm \Sigma \left(\bm \Psi^{-1}\right)'$ via Cholesky, where $\bm \Psi^{-1}$ is a lower unit triangular matrix and $\bm \Sigma = \text{diag}(\sigma^2_{1}, \dots, \sigma^2_{D})$. 

Specifically, on the coefficients of the VAR in Eq. (\ref{eq:state-factor}), we use a horseshoe \citep[HS,][]{carvalho2010horseshoe} prior to regularize the individual parameters in $\bm{A}$. Let $a_{dj}$ denote the $(d,j)^\textsuperscript{th}$ element in $\bm A$ and $\bm a_{d}$ denote the $d^\textsuperscript{th}$ row in $\bm A$. This prior amounts to $a_{dj} \sim \mathcal{N}(0, \varpi^2_{j, \bm a_{d}} \varphi^2_{\bm a_{d}})$, where $\varpi^2_{j, \bm a_{d}}$ refers to a local shrinkage scaling associated with the $(d,j)^\textsuperscript{th}$ element, and $\varphi^2_{\bm a_{d}}$ is a global (factor-specific) shrinkage parameter common to all elements in $\bm a_{d}$. Both hyperparameters are then specified to follow a Half-Cauchy distribution: $\varpi_{j, \bm a_{d}} \sim \mathcal{C}^+(0,1)$ and $\varphi_{\bm a_{d}} \sim \mathcal{C}^+(0,1)$. While the global shrinkage parameter typically pushes all elements in $\bm a_{d}$ strongly towards zero, the local shrinkage scalings allow for element-specific adjustments in the degree of shrinkage. Since both hyperparameters are updated in a data-driven manner, this makes the horseshoe prior particularly powerful and attractive for any VAR setting (small or large), as it provides an adaptive, tuning-free prior specification. 

We also use a HS prior on the contemporaneous relationships. For this, we recast the reduced form VAR in Eq. (\ref{eq:state-factor}) into its structural form \citep[see][]{carriero2022corrigendum}: $\bm \Psi \bm \varepsilon_t = \tilde{\bm \varepsilon}_t,$ $\tilde{\bm \varepsilon}_t \sim \mathcal{N}(\bm 0_D, \bm \Sigma)$, with $\bm{\varepsilon}_{t} = \bm{f}_t - \sum_{p=1}^{P} \bm{A}_{p} \bm{f}_{t-p}$. On the free elements in $\bm{\Psi}$ (the inverse of $\bm{\Psi}^{-1}$, which collects the structural contemporaneous relationships), we specify $\psi_{dj} \sim \mathcal{N}(0, \varpi^2_{j, \bm \psi_{d}} \varphi^2)$, for $d \geq 2$ and $d > j$. Here, $\psi_{dj}$ denotes the $(d,j)^{\text{th}}$ element in $\bm \Psi$ and $\varpi^2_{j, \bm \psi_{d}}$ the associated local shrinkage scaling (with $\bm \psi_d$ indicating the $d^{\text{th}}$ row of $\bm \Psi$) and $\varphi^2$ a global scaling parameter common to all free elements in $\bm{\Psi}$. Analogously to the VAR coefficients above, we define $\varpi_{j, \bm \psi_{d}} \sim \mathcal{C}^+(0,1)$ and $\varphi \sim \mathcal{C}^+(0,1)$. 

For the structural error variances in the state equation, we specify either inverse Gamma priors $\sigma^2_d \sim \mathcal{G}^{-1}\left(\nu_{\sigma}, S_{\sigma}\right)$ or, when using a stochastic volatility (SV) specification, we assume that the log-volatilities follow an AR(1) process: $\log \sigma^2_{dt} = \varsigma_{dt} =\mu_{\varsigma_d} + \rho_{\varsigma_d} \varsigma_{dt-1} + \eta_{\varsigma_d}$, with $\eta_{\varsigma_d} \sim \mathcal{N}(0, \sigma^2_{\varsigma_d})$. The latter has been found to be important in macroeconomic applications. Therefore we consider this straightforward extension in our empirical application(s) as well. This modification does not alter the general structure of the model. The only change is that the constant $\bm \Sigma$ is replaced by a time-varying $\bm \Sigma_t$, featuring a $t$ sub-index. For the prior distributions of the SV parameters, we follow \cite{kastner2014ancillarity} and specify a weakly informative Gaussian prior on $\mu_{\varsigma_d}$, a Beta prior on the transformed autoregressive parameter $( \rho_{\varsigma_d}+1)/2$ and a Gamma shrinkage prior on the state innovation variance $\sigma^2_{\varsigma_d}$, which is supposed to push the model towards homoskedasticity.\footnote{For details, see \cite{kastner2014ancillarity}.}

\subsection{Posterior sampling via MCMC}\label{app:MCMC}
In this Section we provide the full details on the three main steps/blocks of our Markov chain Monte Carlo (MCMC) algorithm.

\begin{enumerate}[leftmargin =*]
\item \textbf{Sample common latent factors.} We use a Particle Gibbs with Ancestor Sampling (PGAS) to update $\{\bm f_t\}_{t = 1}^T$ from its conditional posterior distribution \citep{Andrieu2010, lindsten14a}. \cite{Andrieu2010} show how a version of the particle filter, conditioned on a fixed trajectory for one of the particles can be used to produce draws that result in a Markov kernel with a target distribution that is invariant. However, the usual problem of path degeneracy in the particle filter can result in poor mixing in the original version of particle Gibbs sampler. Recent development suggest that small modifications of this algorithm can largely alleviate this problem. In particular, \cite{lindsten14a} propose the addition of a step that involves sampling the ``ancestors" or indices associated with the particle that is being conditioned on. They show that this results in a substantial improvement in the mixing of the algorithm even with few particles. As explained in \cite{lindsten14a}, ancestor sampling breaks the reference path into segments allowing the particle system to collapse onto a new higher probability path. In the absence of ancestor sampling the particle system tends to collapse to the reference path. Therefore, we use the following particle Gibbs with ancestor sampling step. 

We outline our nonlinear state space model, which we aim to sample from, in Eqs. (\ref{eq:obsapprox}) and (\ref{eq:state-factor}). Eq. (\ref{eq:obsapprox}) is the measurement equation after applying the reduced-rank approximation, and Eq. (\ref{eq:state-factor}) is the state equation, which follows a VAR($P$) process. In the following, the VAR($P$) in Eq. (\ref{eq:state-factor}) can be rewritten more compactly in its companion form:
\begin{equation*}
\tilde{\bm f}_{t}=\bm B \tilde{\bm f}_{t-1}+ \bm e_{t}, \bm e_{t} \sim  \mathcal{N}(0,\tilde{\bm Q}).
\end{equation*}
Here, $\tilde{\bm f}_{t}= (\bm f_t', \bm f_{t-1}', \dots, \bm f_{t-(P-1)}')'$, $\bm B$ denotes the VAR parameters, $
\bm e_{t}$ the errors, $\tilde{\bm Q}$ the variance-covariance matrix in companion form. 

Let $\tilde{\bm f}_{t}^{(s-1)}$, for $t = 1, 2, \dots, T$, denote the fixed trajectory obtained in the previous draw of our Gibbs sampler, corresponding to the $(s - 1)^{\text{th}}$ iteration. Collect all the remaining parameters in the $\bm \Theta $ and denote the number of particles by $H$. To obtain the $s^{\text{th}}$ draw, the conditional particle filter with ancestor sampling proceeds in the following steps. 

For period $t = 1$, we have:
\begin{enumerate}[leftmargin =*, label=(\roman*)]
\item For particles $h=1,2,\dots, (H-1)$, draw $\tilde{\bm f}_{1}^{(h)}|\tilde{\bm f}_{0}^{(h)}$,$\bm \Theta$  and set the $H^{\text{th}}$ particle to the previously accepted draw, $\tilde{\bm f}_{1}^{(H)}=\tilde{\bm f}_{1}^{\left(s-1\right)}$.
\item Compute the normalized weights $\zeta_{1}^{(h)}=\frac{\exp \left(\tilde{\zeta}_{1}^{(h)}\right)}{\sum_{k=1}^{H} \exp \left( \tilde{\zeta}_{1}^{(k)} \right)}$, where $\tilde{\zeta}_{1}^{(h)}$ denotes the conditional log-likelihood defined as:
\begin{equation*}
\log \mathcal{L}\left(\tilde{\bm f}_{1}^{(h)}\right)  = -\frac{1}{2} \log (2 \pi) - \frac{1}{2} \log \left(\det \bm R \right) - \frac{1}{2} \left( \bm v^{(h)}_{1} \bm R^{-1} \bm v^{(h)'}_{1}\right). 
\end{equation*}
Here, $\bm v_{1}^{(h)}$ is computed by simply rearranging the measurement equation in Eq. (\ref{eq:obsapprox}) and plugging in $\bm f_{1}^{(h)}$: $\bm{v}^{(h)}_1 = \bm{y}_1 - \bm{C}\bm{\Phi}(\bm{f}^{(h)}_1)$.
\end{enumerate}
For the subsequent periods $t = 2, \dots, T$ we have: 
\begin{enumerate}[leftmargin =*, label=(\roman*)]
\item Resample $\tilde{\bm f}_{t-1}^{(h)}$ for $h=1,\dots, (H-1)$ using indices $\iota_{t}^{(h)}$ with $\Pr \left(\iota_{t}^{(h)}=h\right) \propto $ $\zeta_{t-1}^{(h)}$.
\item Draw $\tilde{\bm f} _{t}^{(h)}|\tilde{\bm f} _{t-1}^{(\iota_{t}^{(h)})}$,$\bm \Theta$ for $h=1, \dots, (H-1)$ using the state/factor dynamics of the model: 
\begin{equation*}
\tilde{\bm f} _{t}^{(h)}=\bm B \tilde{\bm f} _{t-1}^{(\iota_{t}^{(h)})} + \bm e_{t}, \quad \bm e_{t} \sim  \mathcal{N}(0,\tilde{\bm Q}).
\end{equation*}
Note that $\tilde{\bm f}_{t-1}^{(\iota_{t}^{(h)})}$ denotes the resampled particles from step (i) above.
\item Set the $H^{\text{th}}$ particle to the previously accepted draw $\tilde{\bm f} _{t}^{(H)}=\tilde{\bm f} _{t}^{\left(s-1\right) }$.
\item Sample $\iota_{t}^{(H)}$ with probability defined by weights:
\begin{equation*}
\Pr \left(\iota_{t}^{(H)} = H \right) \propto \zeta_{t-1}^{(h)} \prod_{\tau = t}^{t-1+\tau_1} g(\bm y_{\tau}|\tilde{\bm f}^{(h)}_{1:t-1},\tilde{\bm f}^{(s-1)}_{t:\tau})f(\tilde{\bm f}^{(s-1)}_{\tau}|\tilde{\bm f}^{(h)}_{1:t-1},\tilde{\bm f}^{(s-1)}_{t:\tau-1}).
\end{equation*}
As discussed in \cite{lindsten14a} (see Eq. (23) and Section 7.2 in that paper), this formula is an approximation that can be used when the transition density is degenerate, as in our case. We set the number of lags in this approximation to $\tau_1 = 5$ in our application.  
\item Update the weights $\zeta_{t}^{(h)}=\frac{\exp \left(\tilde{\zeta}_{t}^{(h)}\right)}{\sum_{k=1}^{H} \exp \left( \tilde{\zeta}_{t}^{(k)} \right)}$, where $\tilde{\zeta}_{t}^{(h)}$ denotes the conditional log-likelihood.
\end{enumerate}

Sample $\tilde{\bm f}_{t}^{\left(s\right) }$ with $\Pr \left( \tilde{\bm f}_{t}^{\left( s\right) }=\tilde{\bm f} _{t}^{\left( h\right) }\right) \propto $ $\zeta_{T}^{(h)}$ to obtain a draw from the conditional posterior distribution.

\item \textbf{Sample unknown parameters in the measurement equation.}
Most of the parameters can be sampled on an equation-by-equation basis (i.e., independently over $i = 1, \dots, N$) using well-known conditional posterior distributions. The exception are the GP hyperparameters, which require an equation-specific Metropolis-Hastings (MH) step. Let $\star$ denote conditioning on everything else.
\begin{enumerate}[leftmargin =*, label=(\roman*)]
    \item \textbf{Variance of the idiosyncratic components $r_{i}$:} Using a conditionally conjugate inverse Gamma prior results in an inverse Gamma conditional posterior of the form: 
    \begin{equation}\label{eq:Rpost}
    r_i|\star \sim \mathcal{G}^{-1}\left(T+ \nu_r, S_r + \frac{\sum_{t=1}^{T} v_{it}^2}{2}\right),  
    \end{equation}
    with $v_{it} = y_{it} - \sum_{m=1}^{M} \phi_m(\bm{f}_t)c_{im}$.
    \item \textbf{Coefficients in the observation equation $\bm c_i = (c_{i1}, \dots, c_{iM})'$:} As shown in \citet{svensson2016computationally}, the reduced form approximation outlined in Section \ref{subsec:apprxGP} implies a specific form for the conditional posterior of the coefficients in the observation equation. The conditional posterior is multivariate Gaussian and takes the form:
    \begin{equation}\label{eq:Cpost}
    \bm c_i|\star \sim \mathcal{N}\left(\overline{\bm c}_{i}, \overline{\bm V}_{\bm c_{i}}\right),  
    \end{equation}
    with 
    \begin{equation*}
    \begin{aligned}
    \overline{\bm V}_{\bm c_{i}} =& \left( \frac{\bm \Phi(\bm F)' \bm \Phi(\bm F)}{r_i} + \underline{\bm V}_{\bm c_{i}}^{-1} \right)^{-1}, \\
    \overline{\bm c}_{i} =& \overline{\bm V}_{\bm c_{i}} \left(\frac{\bm \Phi(\bm F)'\bm y_i}{r_i} \right),  
    \end{aligned}
    \end{equation*}
    where $\underline{\bm V}_{\bm c_{i}} = \text{diag}\left(\mathcal{S}_i (\sqrt{\bm \lambda_1}), \dots, \mathcal{S}_i(\sqrt{\lambda_M}) \right)$ and $\bm F = (\bm f_1, \dots, \bm f_T)'$.
    \item \textbf{Kernel parameters $\bm \theta_{i}$:} To draw $\bm \theta_{i}$, we use an MH step with a random walk proposal. This amounts to proposing a candidate from
    \begin{equation}\label{eq:MHprop}
    \bm \theta_{i}^{\ast} = \exp \left(\bm s_{\theta_i} \right) \bm \theta_{i}^{(s-1)},
    \end{equation}
    with $\bm s_{\theta_i}$ being sample from a multivariate Gaussian distribution with zero mean and proposal diagonal variance-covariance matrix, set in such a way that the acceptance rate indicates good mixing of this MH step. The acceptance probability is computed as: 
    \begin{equation}\label{eq:MHaccept}
    \min \left(\left[\log p\left(\bm \theta_{i}^{\ast}|\bm y_i, \star\right) + \log q\left(\bm \theta_{i}^{(s-1)}|\bm \theta_{i}^{\ast}\right) - \log p\left(\bm \theta_{i}^{(s-1)}|\bm y_i, \star \right) - \log q\left(\bm \theta_{i}^{\ast}|\bm \theta_{i}^{(s-1)}\right)\right],0 \right).
    \end{equation}
    Here, we evaluate the conditional log-posterior, $\log p\left(\bm \theta_{i}|\bm y_i, \star\right)$, for both the previously accepted draw $\bm \theta_{i}^{(s-1)}$ and a candidate value $\bm \theta_{i}^{\ast}$ drawn from the proposal. The conditional log-posterior distribution is straightforward to evaluate and given by:  $\log p\left(\bm \theta_{i}^{\ast}|\bm y_i, \star\right) =  \log \mathcal{L}\left(\bm \theta_{i} \right) +  \log p(\bm \theta_i)$, with $\log \mathcal{L}\left(\bm \theta_{i} \right)$ denoting the log-likelihood and $\log p(\bm \theta_i)$ the log-prior distribution.  The conditional log-likelihood is defined as:
    \begin{equation*}
    \log \mathcal{L}\left(\bm \theta_{i} \right)  = -\frac{T}{2} \log (2 \pi)-\frac{T}{2} \log (r_i(\bm \theta_i)) - \frac{\sum_{t=1}^{T} v_{it}(\bm \theta_{i})^2}{2 r_i(\bm \theta_i)},  
    \end{equation*}
    with both $r_i(\bm \theta_i)$ and $v_{it}(\bm \theta_{i})$ being a function of the hyperparameters $\bm \theta_i$.
    The log-prior distribution, $\log p(\bm \theta_i)$ is given by the Gamma distributions outlined in Section \ref{subsec:prior} and a function of the respective hyperparameters $\nu_\xi$, $\nu_\ell$, $S_\xi$, and $S_\ell$. 
\end{enumerate}

\item \textbf{Sample unknown parameters in the state equation.} Conditional on knowing $\bm f_t$, we sample the parameter in the state equation using the algorithm proposed in \cite{carriero2022corrigendum}. 
\begin{enumerate}[leftmargin =*, label=(\roman*)]
\item The contemporaneous relationships among elements in $\bm f_t$ (the state equation covariances) can be sampled from conditional (multivariate) Normal posterior distributions. Let $\bm{\varepsilon}_{t} = \bm{f}_t - \sum_{p=1}^{P} \bm{A}_{p} \bm{f}_{t-p}$ and decompose $\bm Q = \bm \Psi^{-1} \bm \Sigma \left(\bm \Psi^{-1}\right)'$ via Cholesky, where $\bm \Psi^{-1}$ is a lower unit triangular matrix and $\bm \Sigma = \text{diag}(\sigma^2_{1}, \dots, \sigma^2_{D})$. In what follows, we can rewrite the system as: 
\begin{equation*}
\bm \Psi \bm \varepsilon_t = \tilde{\bm \varepsilon}_t, \quad \tilde{\bm \varepsilon}_t \sim \mathcal{N}(\bm 0_D, \bm \Sigma).
\end{equation*}
Using the lower triangular structure of the Cholesky factor, the $d^{\text{th}}$ (for $d = 2, \dots, D$) equation resembles a standard regression model:
\begin{equation*}
\varepsilon_{dt} = - \sum_{j = 1}^{d-1} \psi_{dj} \varepsilon_{jt} + \tilde{\varepsilon}_{dt}, \quad \tilde{\varepsilon}_{dt} \sim \mathcal{N}(0, \sigma^2_{d}),
\end{equation*}
with $\psi_{dj}$ denoting the $(d,j)^{\text{th}}$ element in $\bm \Psi$. To write such a regression model using full data matrices, let $\bm \psi_d$ denote all free $d-1$ elements in the $d^{\text{th}}$ row of $\bm \Psi$, $\bm \varepsilon_d = (\varepsilon_{d1}, \dots, \varepsilon_{dT})'$, $\tilde{\bm \varepsilon}_d = (\tilde{\varepsilon}_{d1}, \dots, \tilde{\varepsilon}_{dT})'$, and define $\bm \varepsilon_{1:(d-1)} = (\bm \varepsilon_{1}, \dots, \bm \varepsilon_{(d-1)})$ as a $T \times (d-1)$ matrix:
\begin{equation*}
\bm \varepsilon_{d} = - \bm \varepsilon_{1:(d-1)} \bm \psi_{d}  + \tilde{\bm \varepsilon}_{d}, \quad \tilde{\bm \varepsilon}_{d} \sim \mathcal{N}(\bm 0_T, \bm \Xi_d), 
\end{equation*}
with $\bm \Xi_d = \sigma_d^2 \bm I_T$.\footnote{In case we allow for SV, $\sigma^2_{dt}$ varies over time and $\bm \Xi_d = \text{diag}(\sigma^2_{d1}, \dots, \sigma^2_{dT})$.} 
We then draw $\bm \psi_{d}$ (for $d = 2, \dots, D$) from conditional posterior which is a multivariate Gaussian distribution: 
\begin{equation*}
\bm \psi_{d}|\bullet  \sim \mathcal{N}\left(\overline{\bm \psi}_{d}, \overline{\bm V}_{\bm \psi_{d}}\right).
\end{equation*}
Here, $\overline{\bm \psi}_{d}$ refers to the posterior mean and $\overline{\bm V}_{\bm \psi_{d}}$ to the posterior variance, which are given by: 
\begin{equation*}
\begin{aligned}
\overline{\bm V}_{\bm \psi_{d}} =& \left(\bm \varepsilon_{1:(d-1)}'\bm \Xi_d^{-1}  \bm \varepsilon_{1:(d-1)} + \underline{\bm V}_{\bm \psi_{d}}^{-1} \right)^{-1}, \\
\overline{\bm \psi}_{d}=& -\overline{\bm V}_{\bm \psi_{d}} \left(\bm \varepsilon_{1:(d-1)}'\bm \Xi_d^{-1} \bm \varepsilon_{d} \right).
\end{aligned}
\end{equation*}
$\underline{\bm V}_{\bm \psi_{d}}$ denotes a $(d-1) \times (d-1)$ diagonal prior variance-covariance matrix, with the prior variances on the main diagonal $\underline{\bm v}_{\bm \psi_{d}}$ being specific to $\bm \psi_{d}$. We impose a HS shrinkage prior on these elements. Following \cite{makalic2015simple}, this implies that the $j^{\text{th}}$ element in $\underline{\bm v}_{\bm \psi_{d}}$ can be defined as $\underline{v}_{j, \bm \psi_{d}} = \varpi^2_{j, \bm \psi_{d}} \varphi^2$, where $\varpi_{j, \bm \psi_{d}}^2$, for $j = 1, \dots, (d-1)$, refers to local scales and $\varphi^2$ is a single global scale. The global and local scale parameters associated with the HS prior can then be sampled from independent inverse Gamma distributions: 
\begin{equation*}
\begin{aligned}
\varpi_{j, \bm \psi_{d}}^2|\star \sim& \mathcal{G}^{-1}\left(1, \frac{1}{\varrho_{j, \bm \psi_{d}}}+\frac{\psi_{dj}^2}{2\varphi^2}\right), \\
\varphi^2|\star \sim& \mathcal{G}^{-1}\left(\frac{D(D-1)+2}{4}, \frac{1}{\vartheta}+ \sum_{d=1}^{D} \left(\sum_{j=1}^{d-1}\frac{\psi_{dj}^2}{2 \varpi_{j, \bm \psi_{d}}^2}\right) \right),\\
\varrho_{j, \bm \psi_{d}}|\star \sim &  \mathcal{G}^{-1}(1, 1 + 1/\varpi^2_{j, \bm \psi_{d}}), \\
\vartheta|\star \sim &  \mathcal{G}^{-1}(1, 1 + 1/\varphi^2), 
\end{aligned}
\end{equation*}
with $\varrho_{j, \bm \psi_{d}}$ and $\vartheta$ denoting auxiliary variables.

\item To update the state innovation error variances, we can:
\begin{itemize}
\item either sample $\{\sigma^2_d\}_{d = 1}^{D}$ from an inverse Gamma distributions: 
\begin{equation*}
\sigma^2_d|\star \sim \mathcal{G}^{-1}(T/2 + 3, \tilde{\bm \varepsilon}_d'\tilde{\bm \varepsilon}_d/2 + 0.3),
\end{equation*}
\item or sample log-volatilities $\{\log \sigma^2_{dt}\}_{d = 1}^{D}$ using the algorithm proposed in \cite{kastner2014ancillarity}.
\end{itemize}
\item To sample the reduced form VAR coefficients $\bm A = (\bm A_1, \dots, \bm A_P)$, which is a $D \times K$-matrix ($K = DP$), on a factor-by-factor basis, we use the updating step proposed in \cite{carriero2022corrigendum}. Note Eq. (\ref{eq:state-factor}) can be written as:  
$\bm f_t = \bm A \bm x_t + \bm{\varepsilon}_{t}$, with $\bm x_t = (\bm f_{t-1}', \dots, \bm f_{t-P}')'$. Let $\bm a_d$ denote the $d^\textsuperscript{th}$ row of $\bm A$ and $\bm A_{d = 0} = (\bm a_1, \dots, \bm a_{d-1}, \bm 0, \bm a_{d-1}, \dots, \bm a_D)'$ being an $D \times K$ matrix, in which we zero out the respective $d^\textsuperscript{th}$ row in $\bm A$. We can manipulate Eq. (\ref{eq:state-factor}) to isolate $\bm a_d$:
\begin{equation*}
\bm \Psi (\bm f_t - \bm A_{d = 0}\bm x_t) =  (\tilde{\bm \psi}_{d} \otimes \bm x_t') \bm a_d +  \tilde{\bm \varepsilon}_t, \quad \tilde{\bm \varepsilon}_t \sim \mathcal{N}(\bm 0_D, \bm \Sigma), 
\end{equation*}
with $\tilde{\bm \psi}_{d}$ denoting the $d^\textsuperscript{th}$ column in $\bm \Psi$. Using the full data matrices, this takes the form of a standard regression model: 
\begin{equation*}
\ddot{\bm y}_{d} =  \ddot{\bm X}_d \bm a_d +  \ddot{\bm \varepsilon}_d, \quad \ddot{\bm \varepsilon}_d \sim \mathcal{N}(\bm 0_{TD}, \ddot{\bm \Sigma}).
\end{equation*}
Here, $\ddot{\bm y}_{d} = (\ddot{\bm y}_{d1}', \dots, \ddot{\bm y}_{dT}')'$ is a $DT \times 1$-vector with the $t^\textsuperscript{th}$ element given by $\ddot{\bm y}_{dt} = \bm \Psi (\bm f_t - \bm A_{d = 0}\bm x_t)$, $\ddot{\bm X}_d = (\bm X \otimes \tilde{\bm \psi}_{d})$ is a $DT \times K$-matrix with with $\bm X$ collecting $\bm x_t'$ on the $t^\textsuperscript{th}$ row, and $\ddot{\bm \Sigma} = \bm I_T \otimes \bm \Sigma$ is a $DT \times DT$ diagonal matrix (since $\bm \Sigma$ is diagonal).\footnote{$\ddot{\bm \Sigma} = \bm I_T \otimes \bm \Sigma$ in case of homoskedastic VAR errors. If we allow for heteroskedastic errors, we have $\ddot{\bm \Sigma} = \text{diag}(\sigma_{11}, \dots, \sigma_{D1}, \dots,  \sigma_{1T},\dots, \sigma_{DT})$.} Note both $\ddot{\bm y}_{d}$ and $\ddot{\bm X}_d$ are factor-specific and therefore feature a $d$ subscript.
For $d = 1, \dots, D$, we can then draw $\bm a_{d}$ from a multivariate Gaussian conditional posterior distribution:
\begin{equation*}
\bm a_{d}|\bullet  \sim \mathcal{N}\left(\overline{\bm a}_{d}, \overline{\bm V}_{\bm a_{d}}\right),
\end{equation*}
with $\overline{\bm a}_{d}$ referring to the posterior mean and $\overline{\bm V}_{\bm a_{d}}$ to the posterior variance. The two posterior moments are given by: 
\begin{equation*}
\begin{aligned}
\overline{\bm V}_{\bm a_{d}} =& \left(\ddot{\bm X}_d'\ddot{\bm \Sigma}^{-1} \ddot{\bm X}_d + \underline{\bm V}_{\bm a_{d}}^{-1} \right)^{-1}, \\
\overline{\bm a}_{d}=& \overline{\bm V}_{\bm a_{d}} \left(\ddot{\bm X}_d' \ddot{\bm \Sigma}^{-1}\ddot{\bm y}_{d} \right).
\end{aligned}
\end{equation*}
$\underline{\bm V}_{\bm a_{d}}$ is a $K \times K$ diagonal prior variance-covariance matrix, with the prior variances on the main diagonal $\underline{\bm v}_{\bm a_{d}}$ being specific to $\bm a_{d}$. Similarly, to the contemporaneous relationships, we use a HS prior for $\bm a_{d}$. This implies $\underline{v}_{j, \bm a_{d}} = \varpi^2_{j, \bm a_{d}} \varphi^2_{\bm a_{d}}$, where $\varpi^2_{j, \bm a_{d}}$, for $j = 1, \dots, K$, are local scales and $\varphi^2_{\bm a_{d}}$ is an factor-specific global scale. We have the following independent inverse Gamma conditional posterior distributions: 
\begin{equation*}
\begin{aligned}
\varpi_{j, \bm a_{d}}^2|\star \sim& \mathcal{G}^{-1}\left(1, \frac{1}{\varrho_{j, \bm a_{d}}}+\frac{a_{dj}^2}{2\varphi^2_{\bm a_{d}}}\right), \\
\varphi^2_{\bm a_{d}}|\star \sim& \mathcal{G}^{-1}\left(\frac{K+1}{2}, \frac{1}{\vartheta_{\bm a_{d}}}+ \sum_{j=1}^{K}\frac{a_{dj}^2}{2 \varpi_{j, \bm a_{d}}^2} \right),\\
\varrho_{j, \bm a_{d}}|\star \sim &  \mathcal{G}^{-1}(1, 1 + 1/\varpi^2_{j, \bm a_{d}}), \\
\vartheta_{\bm a_{d}}|\star \sim &  \mathcal{G}^{-1}(1, 1 + 1/\varphi_{\bm a_{d}}^2), 
\end{aligned}
\end{equation*}
with $\varrho_{j, \bm a_{d}}$ and $\vartheta_{\bm a_{d}}$ denoting auxiliary variables.

\end{enumerate}
\end{enumerate}

\clearpage

\begin{landscape}
\section{Data Appendix}\label{app:data}

\tiny 
\begin{ThreePartTable}
\begin{longtable}{llr}
\caption{List of FRED-QD variables used in the forecasting exercise.} \label{tab:fred} \\
\toprule
\textbf{\texttt{FRED MNEMONIC}} & \textbf{\texttt{Description}} & \textbf{\texttt{Transformation}} \\ 
\midrule
\endfirsthead

\multicolumn{3}{c}%
{{\bfseries \tablename\ \thetable{} -- continued from previous page}} \\
\toprule
\textbf{\texttt{FRED MNEMONIC}} & \textbf{\texttt{Description}} & \textbf{\texttt{Transformation}} \\ 
\midrule
\endhead

\midrule \multicolumn{3}{r}{{Transformation codes: (\texttt{2}) $\Delta x_t$, (\texttt{5}) $\Delta \log (x_t)$, (\texttt{6}) $\Delta^2 \log(x_t)$; continued on next page}} \\
\endfoot
\bottomrule \multicolumn{3}{r}{{Transformation codes: (\texttt{2}) $\Delta x_t$, (\texttt{5}) $\Delta \log (x_t)$, (\texttt{6}) $\Delta^2 \log(x_t)$}} \\
\endlastfoot

GDPC1    & Real Gross Domestic Product      & (\texttt{5}) \\ 
PAYEMS   & All Employees: Total Nonfarm     & (\texttt{5}) \\ 
CPIAUCSL & Consumer Price Index: All Items  & (\texttt{6}) \\ 
FEDFUNDS & Effective Federal Funds Rate     & (\texttt{2}) \\ 
\midrule
PCDGx    & Real Personal Consumption Expenditures: Durable Goods & (\texttt{5}) \\ 
PCESVx   & Real Personal Consumption Expenditures: Services  & (\texttt{5}) \\ 
PCNDx    & Real Personal Consumption Expenditures: Nondurable Goods & (\texttt{5}) \\ 
Y033RC1Q027SBEAx & Real Gross Private Domestic Investment: Fixed Investment (Nonresidential) & (\texttt{5}) \\ 
PNFIx            & Real Private Fixed Investment: Nonresidential & (\texttt{5}) \\ 
PRFIx            & Real Private Fixed Investment: Residential & (\texttt{5}) \\ 
A014RE1Q156NBEA  & Shares of Gross Domestic Product: Gross Private Domestic Investment (Change in Private Inventories) & \texttt{none} \\ 
A823RL1Q225SBEA  & Real Government Consumption Expenditures and Gross Investment (Federal) & \texttt{none} \\ 
FGRECPTx         & Real Federal Government Current Receipts & (\texttt{5}) \\ 
SLCEx            & Real Government State and Local Consumption Expenditures & (\texttt{5}) \\ 
EXPGSC1          & Real Exports of Goods \& Services & (\texttt{5}) \\ 
IMPGSC1          & Real Imports of Goods \& Services & (\texttt{5}) \\ 
IPDMAT           & Industrial Production: Durable Materials & (\texttt{5}) \\ 
IPNMAT           & Industrial Production: Nondurable Materials & (\texttt{5}) \\ 
IPDCONGD         & Industrial Production: Durable Consumer Goods  & (\texttt{5}) \\ 
IPB51110SQ       & Industrial Production: Durable Goods -- Automotive Products  & (\texttt{5}) \\ 
IPNCONGD         & Industrial Production: Nondurable Consumer Goods & (\texttt{5}) \\ 
IPBUSEQ          & Industrial Production: Business Equipment & (\texttt{5}) \\ 
IPB51220SQ       & Industrial Production: Consumer Energy Products & (\texttt{5}) \\ 
CUMFNS           & Capacity Utilization: Manufacturing (SIC)  & \texttt{none} \\ 
DMANEMP          & All Employees: Durable Goods & (\texttt{5}) \\ 
USCONS           & All Employees: Construction  & (\texttt{5}) \\ 
USEHS            & All Employees: Education \& Health Services & (\texttt{5}) \\ 
USFIRE           & All Employees: Financial Activities & (\texttt{5}) \\ 
USINFO           & All Employees: Information Services & (\texttt{5}) \\ 
USPBS            & All Employees: Professional \& Business Services & (\texttt{5}) \\ 
USLAH            & All Employees: Leisure \& Hospitality  & (\texttt{5}) \\ 
USSERV           & All Employees: Other Services & (\texttt{5}) \\ 
USMINE           & All Employees: Mining and Logging & (\texttt{5}) \\ 
USTPU            & All Employees: Trade, Transportation \& Utilities  & (\texttt{5}) \\ 
USTRADE          & All Employees: Retail Trade  & (\texttt{5}) \\ 
USWTRADE         & All Employees: Wholesale Trade  & (\texttt{5}) \\ 
CES9091000001    & All Employees: Government -- Federal  & (\texttt{5}) \\ 
CES9092000001    & All Employees: Government -- State Government & (\texttt{5}) \\ 
CES9093000001    & All Employees: Government -- Local Government & (\texttt{5}) \\ 
LNS14000012      & Unemployment Rate -- 16 to 19 Years & (\texttt{2}) \\ 
LNS14000025      & Unemployment Rate -- 20 Years and over, Men & (\texttt{2}) \\ 
LNS14000026      & Unemployment Rate -- 20 Years and over, Women & (\texttt{2}) \\ 
UEMPLT5          & Number of Civilians Unemployed -- Less Than 5 Weeks & (\texttt{5}) \\ 
UEMP5TO14        & Number of Civilians Unemployed for 5 to 14 Weeks  & (\texttt{5}) \\ 
UEMP15T26        & Number of Civilians Unemployed for 15 to 26 Weeks & (\texttt{5}) \\ 
UEMP27OV         & Number of Civilians Unemployed for 27 Weeks and Over & (\texttt{5}) \\ 
LNS12032194      & Employment Level -- Part-Time for Economic Reasons, All Industries & (\texttt{5}) \\ 
AWHMAN           & Average Weekly Hours of Production and Nonsupervisory Employees: Manufacturing & \texttt{none} \\ 
AWHNONAG         & Average Weekly Hours Of Production And Nonsupervisory Employees: Total Private & (\texttt{2}) \\ 
AWOTMAN          & Average Weekly Overtime Hours of Production and Nonsupervisory Employees: Manufacturing & (\texttt{2}) \\ 
PERMIT           & New Private Housing Units Authorized by Building Permits & (\texttt{5}) \\ 
HOUSTMW          & Housing Starts in Midwest Census Region & (\texttt{5}) \\ 
HOUSTNE          & Housing Starts in Northeast Census Region & (\texttt{5}) \\ 
HOUSTS           & Housing Starts in South Census Region  & (\texttt{5}) \\ 
HOUSTW           & Housing Starts in West Census Region  & (\texttt{5}) \\ 
AMDMNOx          & Real Manufacturers' New Orders: Durable Goods & (\texttt{5}) \\ 
AMDMUOx          & Real Value of Manufacturers' Unfilled Orders for Durable Goods Industries  & (\texttt{5}) \\ 
GPDICTPI         & Gross Private Domestic Investment: Chain-type Price Index & (\texttt{6}) \\ 
IPDBS            & Business Sector: Implicit Price Deflator & (\texttt{6}) \\ 
DMOTRG3Q086SBEA  & Personal Consumption Expenditures: Durable Goods -- Motor Vehicles and Parts  & (\texttt{6}) \\ 
DFDHRG3Q086SBEA  & Personal Consumption Expenditures: Durable Goods -- Furnishings and Durable Household Equipment & (\texttt{6}) \\ 
DREQRG3Q086SBEA  & Personal Consumption Expenditures: Durable Goods -- Recreational Goods and Vehicles & (\texttt{6}) \\ 
DODGRG3Q086SBEA  & Personal Consumption Expenditures: Durable Goods -- Other Durable Goods & (\texttt{6}) \\ 
DFXARG3Q086SBEA  & Personal Consumption Expenditures: Nondurable Goods -- Food and Beverages Purchased for Off-Premises Consumption & (\texttt{6}) \\ 
DCLORG3Q086SBEA  & Personal Consumption Expenditures: Nondurable Goods -- Clothing and Footwear & (\texttt{6}) \\ 
DGOERG3Q086SBEA  & Personal Consumption Expenditures: Nondurable Goods -- Gasoline and Other Energy Goods & (\texttt{6}) \\ 
DONGRG3Q086SBEA  & Personal Consumption Expenditures: Nondurable Goods -- Other Nondurable Goods  & (\texttt{6}) \\ 
DHUTRG3Q086SBEA  & Personal Consumption Expenditures: Services -- Housing and Utilities  & (\texttt{6}) \\ 
DHLCRG3Q086SBEA  & Personal Consumption Expenditures: Services -- Health care (chain-type price index) & (\texttt{6}) \\ 
DTRSRG3Q086SBEA  & Personal Consumption Expenditures: Transportation Services & (\texttt{6}) \\ 
DRCARG3Q086SBEA  & Personal Consumption Expenditures: Recreation Services  & (\texttt{6}) \\ 
DFSARG3Q086SBEA  & Personal Consumption Expenditures: Services -- Food Services and Accommodations & (\texttt{6}) \\ 
DIFSRG3Q086SBEA  & Personal Consumption Expenditures: Financial Services and Insurance  & (\texttt{6}) \\ 
DOTSRG3Q086SBEA  & Personal Consumption Expenditures: Other Services  & (\texttt{6}) \\ 
WPSFD49502       & Producer Price Index by Commodity for Final Demand: Personal Consumption Goods (Finished Consumer Goods) & (\texttt{6}) \\ 
WPSFD4111        & Producer Price Index by Commodity for Finished Consumer Foods & (\texttt{6}) \\ 
PPIIDC           & Producer Price Index by Commodity Industrial Commodities  & (\texttt{6}) \\ 
WPSID61          & Producer Price Index by Commodity Intermediate Materials: Supplies \& Components & (\texttt{6}) \\ 
WPU0561          & Producer Price Index by Commodity for Fuels and Related Products and Power: Crude Petroleum (Domestic Production) & (\texttt{5}) \\ 
COMPRNFB         & Nonfarm Business Sector: Real Compensation Per Hour & (\texttt{5}) \\ 
RCPHBS           & Business Sector: Real Compensation Per Hour & (\texttt{5}) \\ 
OPHNFB           & Nonfarm Business Sector: Real Output Per Hour of All Persons & (\texttt{5}) \\ 
ULCNFB           & Nonfarm Business Sector: Unit Labor Cost & (\texttt{5}) \\ 
UNLPNBS          & Nonfarm Business Sector: Unit Nonlabor Payments  & (\texttt{5}) \\ 
TB3MS            & 3-Month Treasury Bill: Secondary Market Rate & (\texttt{2}) \\ 
BAA10YM          & Moody's Seasoned Baa Corporate Bond Yield Relative to Yield on 10-Year Treasury Constant Maturity & \texttt{none} \\ 
TB6M3Mx          & 6-Month Treasury Bill Minus 3-Month Treasury Bill & \texttt{none} \\ 
GS1TB3Mx         & 1-Year Treasury Constant Maturity Minus 3-Month Treasury Bill & \texttt{none} \\ 
GS10TB3Mx        & 10-Year Treasury Constant Maturity Minus 3-Month Treasury Bill & \texttt{none} \\ 
BUSLOANSx        & Real Commercial and Industrial Loans, All Commercial Banks & (\texttt{5}) \\ 
CONSUMERx        & Real Consumer Loans at All Commercial Banks & (\texttt{5}) \\ 
NONREVSLx        & Total Real Nonrevolving Credit Owned and Securitized, Outstanding & (\texttt{5}) \\ 
REALLNx          & Real Real Estate Loans, All Commercial Banks & (\texttt{5}) \\ 
TLBSHNOx         & Real Total Liabilities of Households and Nonprofit Organizations & (\texttt{5}) \\ 
TNWBSHNOx        & Real Net Worth of Households and Nonprofit Organizations  & (\texttt{5}) \\ 
TARESAx          & Real Assets of Households and Nonprofit Organizations excluding Real Estate Assets & (\texttt{5}) \\ 
HNOREMQ027Sx     & Real Real Estate Assets of Households and Nonprofit Organizations & (\texttt{5}) \\ 
TFAABSHNOx       & Real Total Financial Assets of Households and Nonprofit Organizations & (\texttt{5}) \\ 
VIXCLSx          & CBOE Volatility Index & (\texttt{5}) \\ 
EXSZUSx          & Switzerland / US Foreign Exchange Rate & (\texttt{5}) \\ 
EXJPUSx          & Japan / US Foreign Exchange Rate & (\texttt{5}) \\ 
EXUSUKx          & UK / US Foreign Exchange Rate & (\texttt{5}) \\ 
EXCAUSx          & Canada / US Foreign Exchange Rate & (\texttt{5}) \\ 
UMCSENTx         & University of Michigan: Consumer Sentiment & \texttt{none} \\ 
SP500            & S\&P 500 Common Stock Price Index:  Composite & (\texttt{5}) \\ 
\end{longtable}
\end{ThreePartTable}

\end{landscape}

\end{appendices}
\end{document}